\documentclass[useAMS,usenatbib]{mn2e}

\usepackage{graphicx}
\usepackage{natbib}
\usepackage{amsmath}
\usepackage{amssymb}
\usepackage{amsfonts}
\usepackage{xspace}
\usepackage[normalem]{ulem}
\usepackage{color}

\newcommand{\dd}{\mathrm{d}}
\newcommand{\ltsima}{$\; \buildrel < \over \sim \;$}
\newcommand{\lsim}{\lower.5ex\hbox{\ltsima}}
\newcommand{\gtsima}{$\; \buildrel > \over \sim \;$}
\newcommand{\gsim}{\lower.5ex\hbox{\gtsima}}
\newcommand{\eps}{\varepsilon}

\newcommand{\pro}{\mathrm{p}}

\newcommand{\e}{\mathrm{e}}
\newcommand{\inj}{\mathrm{inj}}

\newcommand{\CR}{\mathrm{CR}}

\newcommand{\CMB}{\mathrm{CMB}}
\newcommand{\IC}{\mathrm{IC}}
\newcommand{\mec}{m_\e\,c}
\newcommand{\mug}{\mathrm{\mu G}}
\newcommand{\rvir}{R_{200}}
\newcommand{\mvir}{M_{200}}
\newcommand{\msol}{\rmn{M}_\odot}
\newcommand{\B}{\mathrm{B}}
\newcommand{\pC}{p_\mathrm{Coul}}
\newcommand{\pIC}{p_\mathrm{IC}}
\newcommand{\pCt}{\tilde{p}_\mathrm{Coul}}
\newcommand{\pICt}{\tilde{p}_\mathrm{IC}}
\newcommand{\pf}{p_\mathrm{f}}
\newcommand{\pin}{p_\mathrm{i}}
\newcommand{\CRe}{\mathrm{CRe}}
\newcommand{\tend}{t_\rmn{f}}
\newcommand{\mach}{\mathcal{M}}
\newcommand{\eV}{\mathrm{eV}}
\newcommand{\cm}{\mathrm{cm}}
\newcommand\pinj{p_{\mathrm{inj}}}
\newcommand\ainj{a_{\mathrm{inj}}}
\newcommand\rtot{r}
\newcommand\xinj{x_{\mathrm{inj}}}
\newcommand\eb{\epsilon_\rmn{B}}
\newcommand{\mae}{m_\e}
\newcommand{\alpi}{\alpha_\rmn{inj}}
\newcommand{\alps}{\alpha_\rmn{sub}}
\newcommand{\kev}{\rmn{keV}}
\newcommand\nth{n_{\rmn{th}}}
\newcommand\gamad{\gamma_\rmn{ad}}

\newcommand\pcut{p_{\mathrm{cut}}}

\newcommand{\mpc}{\rmn{Mpc}}
\newcommand{\nus}{\nu_\rmn{s}}

\title{Giant radio relics in galaxy clusters: reacceleration of fossil relativistic electrons?}

\author[A. Pinzke, S.P. Oh and C. Pfrommer] 
  {Anders Pinzke$^1$\thanks{e-mail:apinzke@fysik.su.se (AP); peng@physics.ucsb.edu (PO);pfrommer@h-its.org (CP)}, S. Peng Oh$^1$\footnotemark[1],
    and Christoph Pfrommer$^2$\footnotemark[1]\\
    $^1$Department of Physics, University of California, 
    Santa Barbara, CA 93106-9530, USA\\
    $^2$Heidelberg Institute for Theoretical Studies
  (HITS), Schloss-Wolfsbrunnenweg 35, DE - 69118 Heidelberg, Germany}

\begin{document}
\pagerange{\pageref{firstpage}--\pageref{lastpage}} \pubyear{2011}
\maketitle
\label{firstpage}

\begin{abstract}
Many bright radio relics in the outskirts of galaxy clusters have low
inferred Mach numbers, defying expectations from shock acceleration
theory and heliospheric observations that the injection efficiency of
relativistic particles plummets at low Mach numbers. With a suite of
cosmological simulations, we follow the diffusive shock acceleration
as well as radiative and Coulomb cooling of cosmic ray electrons
during the assembly of a cluster. We find a substantial population of
fossil electrons. When reaccelerated at a shock (through diffusive
shock acceleration), they are competitive with direct injection at
strong shocks and overwhelmingly dominate by many orders of magnitude
at weak shocks, $\mach \lsim 3$, which are the vast majority at the
cluster periphery. Their relative importance depends on cooling
physics and is robust to the shock acceleration model used. While the
abundance of fossils can vary by a factor of $\sim 10$, the typical
reaccelerated fossil population has radio brightness in excellent
agreement with observations. Fossil electrons with $1 \lsim \gamma
\lsim 100$ ($10 \lsim \gamma \lsim 10^{4}$) provide the main seeds for
reacceleration at strong (weak) shocks; we show that these are
well-resolved by our simulation. We construct a simple self-similar
analytic model which assumes steady recent injection and cooling. It
agrees well with our simulations, allowing rapid estimates and
physical insight into the shape of the distribution function. We
predict that LOFAR should find many more bright steep-spectrum radio
relics, which are inconsistent with direct injection. A failure to
take fossil cosmic ray electrons into account will lead to erroneous
conclusions about the nature of particle acceleration at weak shocks;
they arise from well-understood physical processes and cannot be
ignored.
\end{abstract}


\section{Introduction}
\label{sect:intro}
Diffuse radio emission in clusters falls into two broad classes:
smooth, centrally located and unpolarized radio halos \citep[][ and
  references therein]{ferrari08}, and elongated, significantly
polarized and steep spectrum radio relics which are seen at cluster
outskirts \citep[][ and references therein]{kempner04}. The radio
halos come in two distinct classes: mini-halos in cool core clusters
and giant halos associated with merging clusters. Similar to the radio
halos, the radio relics are thought to come in distinct classes which
are all associated with merging clusters: fossil radio lobes blown in
the past by active galactic nuclei (AGN), which may have been
re-energized by compression \citep[radio phoenix;][]{ensslin01}, and
those produced by direct diffusive particle acceleration at
accretion/merger shocks \citep[radio
  gischt;][]{ensslin98,miniati01}. Many radio relics have now been
seen \citep[$\sim 50$; for recent compilations,
  see][]{2012A&ARv..20...54F,nuza11} and the number could increase
dramatically with upcoming low frequency radio surveys proposed for
the Low Frequency Array (LOFAR), the Westerbork Synthesis Radio
Telescope (WSRT), and farther in the future, Square Kilometer Array
(SKA). Using numerical simulations combined with a semi-analytic model
for radio emission calibrated to existing number counts,
\citet{nuza11} estimate that LOFAR and WSRT could discover $\sim 2500$
relics and $\sim 900$ relics respectively. The time is therefore ripe
to understand how we can best mine these future surveys.

In this paper, we focus on radio gischt, which trace structure
formation shock waves. They therefore can illuminate the nature of
cosmic accretion/mergers, as well as shock amplification of
large-scale magnetic fields \citep{2008MNRAS.385.1211P,pfrommer08,
  hoeft08, battaglia09, skillman11}. Perhaps even more importantly,
they allow us to probe in detail the efficiency of shock acceleration
in a diffuse, low Mach number ($\mach \sim 2-4$) regime, far different
from the high Mach number regimes probed in our Galaxy and supernova
remnants. Whilst this remains poorly understood, at face value the
observations seem to suggest an electron acceleration efficiency
significantly in excess of naive theoretical expectations. 

A recent spectacular example of radio gischt is CIZA J2242.8+5301, the
`sausage relic' \citep{2010Sci...330..347V}, a large ($\sim 2$ Mpc
long; located $\sim 1.5$ Mpc from the cluster center) double radio
relic system. The post-shock radio spectral index was used to infer
the particle spectral slope and hence the shock compression ratio and
Mach number ($\mach \sim 4.6$), while the decrease in the spectral
index (from $\sim 0.6$ to $\sim 2.0$ across the relic's narrow $\sim
55$ kpc width) toward the cluster center---spectral aging due to
synchrotron and inverse Compton losses---was used to infer magnetic
field strengths of $\sim 5 \, \mu$G. The strong ($\sim 50-60$ per
cent) polarization can be attributed to magnetic field frozen into the
compressed intracluster medium (ICM), which has been aligned parallel
to the shock. The properties of this and similar systems are distinct
from fossil radio plasma, which are smaller, have curved, steeper
spectra (due to aging), and lobe- or torus-like morphology
\citep[e.g.,][]{ensslin02, 2011ApJ...730...22P}. The power-law
spectral index, spectral gradient, and enormous extent clearly support
a diffusive shock acceleration (DSA) origin. While turbulent
acceleration may be a viable mechanism for energizing CRes in radio
halos, it is less plausible for radio relics because of the long
acceleration time scales (exceeding several 100~Myrs). The coincidence
of radio relic emission with X-ray determined shock fronts over scales
of several Mpc imply the injection or reacceleration of CRes on very
short timescales that preclude turbulent reacceleration. Otherwise,
for instance, for an acceleration timescale of $t_{\rm accel} \sim 3
\times 10^{8}$yr and a postshock velocity of $ \sim 1000/4 \sim 250 \,
       {\rm km \, s^{-1}}$, the radio emission and the shock front
       would be separated by $\sim 100$ kpc. In addition, turbulent
       reacceleration models produce curved spectra, which are not
       seen in radio relics.

However, this then presents a puzzle. Cosmological simulations show
that while gas initially undergoes strong shocks (up to $\mach \sim
10^{3}$) accreting onto non-linear structures and filaments, shocks in
the ICM and cluster outskirts are relatively modest ($\mach \sim
1-4$), since the gas has already reached sub-keV temperatures
\citep{2003ApJ...593..599R,2006MNRAS.367..113P,skillman08, vazza09,
  vazza11}. While DSA is efficient in accelerating particles in the
thermal Maxwellian tail at high Mach numbers
\citep{1978MNRAS.182..443B,drury83}, and as confirmed in observations
of supernova remnants \citep{parizot06,reynolds08}, at lower Mach
numbers the efficiency of DSA is known to plummet exponentially.
Indeed, in the test-particle regime where suprathermal particles
undergo acceleration via a thermal leakage process (which compare well
against kinetic DSA simulations) the acceleration efficiency for weak
shocks $\mach \lesssim 3$ is extremely small; the fraction of protons
accelerated is $\sim 10^{-4}-10^{-3}$, and cosmic ray proton (CRp)
pressure is $\lsim 1$ per cent of the shock ram pressure
\citep{kang11}. The acceleration efficiency of electrons at low Mach
numbers is likely to be far smaller still. The injection problem for
thermal electrons is already known to be severe at high Mach numbers,
due to the smaller gyroradius of thermal electrons: the relative
acceleration efficiency of cosmic ray electrons (CRes) is lower by
$\sim 10^{-2}$ as in the Galaxy \citep{schlickeiser02} or even $\sim
10^{-4}$ as in supernova remnants \citep{morlino09}. These relative
efficiencies likely plummets further at low Mach numbers, as is also
suggested by heliospheric observations (see
\S\ref{sec:discussion}). These considerations appear to contradict the
appearance of bright radio relics, and perhaps suggest that our
understanding of DSA at low Mach numbers is incomplete \citep[for
recent progress, see][]{gargate12}.

A possible solution is if there is a pre-existing population of CRes
with gyroradii comparable or larger than that of the shock
thickness. In this case, injection from the thermal pool is no longer
an issue
\citep{ensslin98,markevitch05,giacintucci08,kang11,kang12}. DSA has a
much larger effect on radio emission than adiabatic compression;
including the downstream magnetic field amplification, synchrotron
emission from fossil electrons could be boosted by a factor $\sim
100-1000$ for $\mach \sim 3$ \citep{kang11}. Thus, the luminosity
function of radio gischt will be strongly modified by the presence or
absence of a seed relativistic electron population, whose existence
has never been directly demonstrated. Note that secondary CRe which
arise from hadronic interactions of CRp are not thought to be
significant at these low densities, though for a dissenting view, see
\citet{keshet10}. In this paper, we use our existing
high-resolution SPH simulations of CRps in clusters \citep{pinzke10}
to infer whether structure formation shocks could generate
them\footnote{They could also have a non-gravitational origin, such as
  acceleration by AGNs or SN-driven winds, though the filling factor
  at these large radii is likely to be small and the effect of
  adiabatic cooling during the expansion from the compact interstellar
  medium into the ICM makes their energy fraction likely negligible in
  comparison to those CRes accelerated at structure formation
  shocks.}. Note that DSA operates identically on relativistic
particles of the same rigidity ($R=Pc/Ze$), so the injected proton and
electron spectrum are the same, modulo their relative acceleration
efficiency, which we calibrate off Galactic observations.

\begin{figure}
\begin{center}
{\includegraphics[width=\columnwidth]{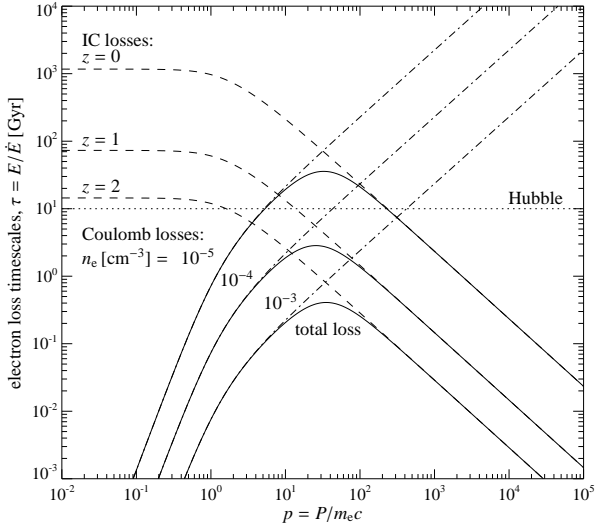}}
\end{center}
\caption{Cooling time of CR electrons for typical conditions in the
  ICM. CRes can be long-lived if they are injected at low density and
  at late times. Electrons with kinetic energy $E\simeq p\, m_\rmn{e}
  c^2 \simeq 10$~MeV should be the most long-lived in cluster
  outskirts ($n_\rmn{e}\sim3\times10^{-5}\rmn{cm}^{-3}$) because this
  energy range maximizes the Coulomb and inverse Compton (IC) cooling
  times.}
\label{fig:timescales}
\end{figure}

The main difference is that unlike protons, relativistic electrons can
still undergo significant Coulomb and inverse Compton losses in the
cluster outskirts. In Fig.~\ref{fig:timescales}, we show the cooling
time of CRes for difference densities and cosmological epochs. CRes of
energy $\sim 10$ GeV that are responsible for $\sim$ GHz emission in a
$\mu$G field have energy loss timescales of $\sim 10^{8}$yr at $z=0$
(and cool even more quickly at higher redshift). However, lower energy
CRe could potentially survive to be reaccelerated since CRes with
momentum $P=(10-100)\,\mec$ have $t_{\rm cool} > t_{\rm H}$ in gas
with number densities $n_{\rm e}\le3 \times 10^{-5} \, \cm^{-3}$
(typical of cluster outskirts) at $z=0$. We evolve the time-dependent
cosmic ray (CR) energy equation to track the evolving distribution
function of CRe electrons under the combined influence of injection
and cooling. We find that the fossil electron population is
substantial, and that reacceleration of this population is competitive
with direct injection for strong shocks, and will overwhelming
dominate radio emission for weak shocks with Mach numbers ${\mathcal
  M} < 4$. Since the latter constitute the vast majority of shocks in
the cluster periphery, reacceleration is a crucial mechanism which
cannot be ignored.

The outline of this paper is as follows. In \S\ref{sect:simulations},
we describe our CR formalism and computational method. In
\S\ref{sect:spectrum_results}, we present results for the fossil
electron spectrum, and an analytic model for understanding its
essential features. In \S\ref{sec:reaccel}, we study how the fossil
population is transformed by a shock, time-resolution effects, and our
principal result, that reacceleration dominates direct injection. In
\S\ref{sect:variations}, we study how the fossil spectrum varies with
the shock acceleration model and between clusters. In
\S\ref{sec:flux}, we discuss observational consequences, including the
brightness of relics as a function of Mach number and the relic
luminosity function. In \S\ref{sec:discussion}, we discuss insights
gained from heliospheric observations on our adopted shock physics as
well as possible limitations of our numerical method. We conclude in
\S\ref{sect:conclusions}. Three appendices also spell out various
technical points about the cooling and injection process.

\section{Fossil Electrons: Method}
\label{sect:simulations}
Our simulations adopt a $\Lambda$CDM cosmology with parameters:
$\Omega_{\rm m}=\Omega_{\rm DM} + \Omega_{\rm b}=0.3, \, \Omega_{\rm
  b}=0.039, \, \Omega_\Lambda=0.7, \, h=0.7, \, n_{\rm s}=1$, and
$\sigma_8=0.9$. The simulations were carried out with an updated and
extended version of the distributed-memory parallel TreeSPH code
GADGET-2 \citep{2005MNRAS.364.1105S,
  2001NewA....6...79S}. Gravitational forces are computed using a
combination of particle-mesh and tree algorithms.  Hydrodynamic forces
are computed with a variant of the smoothed particle hydrodynamics
(SPH) algorithm that conserves energy and entropy where appropriate,
i.e. outside of shocked regions \citep{2002MNRAS.333..649S}.  Our
simulations follow the radiative cooling of the gas, star formation,
supernova feedback, and a photo-ionizing background \citep[details can
  be found in][]{2007MNRAS.378..385P}. We model the CR physics in a
self-consistent way \citep{2006MNRAS.367..113P,2007A&A...473...41E,
  2008A&A...481...33J} and attach a CRp distribution function to each
SPH fluid element. We include adiabatic CRp transport process such as
compression and rarefaction, and a number of physical source and sink
terms which modify the CRp pressure of each CRp population
separately. The main source of injection is diffusive shock
acceleration (DSA) at cosmological structure formation shocks, while
the primary CRp sinks are thermalization by Coulomb interactions, and
catastrophic losses by hadronization. We do not consider CRps injected
by supernovae remnants or AGN feedback; these sources should be
relatively subdominant in the cluster outskirts. In addition, we
neglect the reacceleration of CRp. Firstly, the CRp do not affect
radio emission, and hence are unimportant for our study. Secondly, in
any case, the CRp population barely cools; only adiabatic cooling
(which is an order unity effect) is significant. Reacceleration is
only important in counteracting the effects of cooling; otherwise, the
effect of the predominantly weak shocks which produce reacceleration
on the hard power law of a strong shock is negligible.

\begin{table}
\caption{Cluster sample.}
\begin{tabular}{l l c r r l}
\hline
\hline
Sim.'s & State$^{(1)}$ & $\mvir^{(2)}$ & $\rvir^{(2)}$ & $kT_{\rm 200}^{(3)}$ & $\Delta t, \frac{a_2}{a_1}$$^{(4)}$\\
& & [$\msol$] & [Mpc] & [keV] & \\
\hline
g8a  & CC    & $2.6\times 10^{15}$ &   2.9~~ & 13.1 & 1.127 \\
g1a  & CC    & $1.9\times 10^{15}$ &   2.5~~ & 10.6 & 1.127 \\
g72a & PostM & $1.6\times 10^{15}$ &   2.4~~ & 9.4  & 100 Myr \\
g51  & CC    & $1.5\times 10^{15}$ &   2.4~~ & 9.4  & 1.127 \\
                                                     
g1b  & M     & $5.2\times 10^{14}$ &   1.7~~ & 4.7  & 1.127 \\
g72b & M     & $2.2\times 10^{14}$ &   1.2~~ & 2.4  & 100 Myr \\
g1c  & M     & $2.0\times 10^{14}$ &   1.2~~ & 2.3  & 1.127 \\
g8b  & M     & $1.5\times 10^{14}$ &   1.1~~ & 1.9  & 1.127 \\
g1d  & M     & $1.3\times 10^{14}$ &   1.0~~ & 1.7  & 1.127 \\
                                                     
g676 & CC    & $1.3\times 10^{14}$ &   1.0~~ & 1.7  & 1.049 \\
g914 & CC    & $1.2\times 10^{14}$ &   1.0~~ & 1.6  & 1.049 \\
g1e  & M     & $9.1\times 10^{13}$ &  0.93   & 1.3  & 1.127 \\
g8c  & M     & $8.5\times 10^{13}$ &  0.91   & 1.3  & 1.127 \\
g8d  & PreM  & $7.8\times 10^{13}$ &  0.88   & 1.2  & 1.127 \\
\hline
\end{tabular}  \begin{quote}
 Notes:\\ (1) The dynamical state has been classified through a
 combined criterion invoking a merger tree study and visual inspection
 of the X-ray brightness maps. The labels for the clusters are:
 PreM--pre-merger (sub-cluster already within the virial radius),
 M--merger, PostM--post-merger (slightly elongated X-ray contours,
 weak cool core (CC) region developing), CC--cool core cluster with
 extended cooling region (smooth X-ray profile).  (2) The virial mass
 and radius are related by $M_\Delta(z) = \frac{4}{3} \pi\, \Delta\,
 \rho_\rmn{crit}(z) R_\Delta^3 $, where $\Delta=200$ denotes a
 multiple of the critical overdensity $\rho_\rmn{crit}$.  (3) The
 virial temperature is $kT_{\rm 200} = G \mvir \, \mu\, m_\pro / (2
 \rvir)$, where $\mu$ is the mean molecular weight.  (4) Time
 difference between output snapshots; in units of Myr for g72, and
 remaining clusters show the ratio of the cosmological scale factor,
 $a$, between two snapshots (roughly corresponding to a time interval
 $\Delta t \sim 1$Gyr).
\end{quote}
\label{tab:cluster_sample}
\end{table}

In this paper, we post-process previous cosmological simulations of 14
galaxy clusters \citep{pinzke10}, adapting them to study radio
relics. The properties of the cluster sample is listed in Table
\ref{tab:cluster_sample}. In our previous work, CRps were accelerated
through DSA on the fly in the simulations, and the CRp distribution
was written out into snapshots. We now use this information to
identify shocks and inject CRes. In particular, we:
\begin{itemize}
\item identify all SPH particles that have undergone a shock by
  comparing the CRp distribution function between snapshots
\item inject CRes according to acceleration scheme (described in \S\ref{sec:models})
\item evolve each injected CRe population to a later time $\tend$
  while accounting for losses (Coulomb, inverse Compton, and
  adiabatic)
\item add up the CRe distribution for all SPH particles at time
  $\tend$: this constitutes our fossil spectrum. 
\item reaccelerate CRe distribution function at time $\tend$ using
  typical values for a merging shock in cluster outskirts
\item calculate the radio synchrotron emission from radio relics
\end{itemize}

In this section, we describe our scheme for deriving the fossil CRe
distribution function at time $t_{\rm f}$. Reacceleration and radio
emission are described in \S\ref{sec:reaccel} and \S\ref{sec:flux}
respectively.

\subsection{Basic CR Formalism}
\label{sec:basics}
The CR electrons and protons are each represented by an isotropic
one-dimensional distribution function,\footnote{The three-dimensional
  distribution function is $f^{(3)}(p)=f(p)/(4\pi\,p^2)$.} which we
assume to be a superposition of power-law spectra, each represented by
\begin{equation}
  \label{eq:f_p}
  f(p) \equiv \frac{\dd^2 N}{\dd p\,\dd V} = 
  C\, p^{-\alpha}\,H(p-\pcut)\,, 
\end{equation}
where $p \equiv P/m\,c$, where $P$ is the momentum, $m$ the mass of
the particle, and $c$ the speed of light. Note that for Lorentz
factors $\gamma \gg 1$, $p \approx \gamma$. Additionally, $\pcut$ is
the momentum cutoff, $\alpha$ the spectral index of the CR power-law
distribution, $C$ the normalization of the distribution function, and
$H(x)$ is the Heaviside step function. The differential CR spectrum
can vary spatially and temporally, but for brevity we suppress this in
our notation.

The number density of a single power-law CR distribution is given by
\begin{equation}
\label{eq:ncr}
n_{\CR} = \int_0^\infty  \dd p\, f(p) =
\frac{C\, \pcut^{1-\alpha}}{\alpha-1}\,
\end{equation}
provided $\alpha >1$. The kinetic energy density of the CR population
is
\begin{eqnarray}
\label{eq:epscr}
&&\eps_\CR = \int_0^\infty  \dd p\, f(p) \,T(p)=\frac{C\,
  m\,c^2}{\alpha-1} \, \times \nonumber \\
&&\left[\frac{1}{2}
\, \B_x \left(
\frac{\alpha-2}{2},\frac{3-\alpha}{2}\right) + \pcut^{1-\alpha}
 \left(\sqrt{1+\pcut^2}-1 \right) \right],\nonumber \\
&&
\end{eqnarray}
where $T(p) = (\sqrt{1+p^2} -1)\, m\,c^2$ is the kinetic energy of a
particle with momentum $p$ and $x = 1/(1+\pcut^2)$. $\B_x(a,b)$
denotes the incomplete Beta-function, and $\alpha>2$ is assumed. The
average CR kinetic energy $T_\CR = \eps_\CR/n_\CR$ is therefore
\begin{equation}
\label{eq:Tcr}
T_\CR = \left[\frac{\pcut^{\alpha-1}}{2} \, \B_x
\left( \frac{\alpha-2}{2},\frac{3-\alpha}{2}\right) + \sqrt{1+\pcut^2}-1
\right]\,{m\,c^2}\,.
\end{equation}

\subsection{CR Injection} 
\subsubsection{Identifying Shocks}
In our cosmological simulations, we model the CR proton (CRp)
distribution function as a superposition of CRp populations, each
determined by equation~\eqref{eq:f_p}, but with a different spectral
index ($\alpha=\{2.1,2.3,2.5,2.7,2.9\}$), momentum cut-off $\pcut$,
and normalization $C_\pro$ (where the subscript denotes the CR proton
population) derived from the simulations. Both $C_{\rm p}=C(x,t)$ and
$p_{\rm cut}=p(x,t)$ are allowed to vary spatially and temporally.

We identify the SPH particles that have experienced a shock by
calculating the change in the adiabatic invariant of the CR
normalization, $C_{0,\pro}=(\rho/\rho_0)^{-(\alpha+2)/3}C_{\pro}$,
between a snapshot at time $t$ and an earlier time $t-\Delta t$:
\begin{equation}
 \label{eq:finj}
 \Delta C_{0,\pro}(t) = C_{0,\pro}(t) - C_{0,\pro}(t-\Delta t)\,.
\end{equation}
Here the time between snapshots is denoted by $\Delta t$. It is shown
in Table~\ref{tab:cluster_sample} for each simulated cluster. We
resolve cooling and injection in two of the simulated clusters (g72a
and g72b) on $\Delta t \sim$100~Myr timescales, while the remaining
clusters in our sample are resolved on longer timescales $\Delta t\sim
1$~Gyr. If the condition $\Delta C_{0,\pro}(t) > 0$ is fulfilled, then
the CRps have experience at least one shock within the time $\Delta
t$, since the cooling time of CRps on the cluster outskirts is
sufficiently long that they are essentially adiabatic.

\subsubsection{Diffusive Shock Acceleration}
\label{sec:DSA}
Here we introduce the basic framework of DSA which we use. In our
discussions $(\rho_{0},u_{0})$ and $(\rho_{2},u_{2})$ refer to
upstream and downstream densities/velocities in the shock rest frame,
respectively. In the thermal leakage model for DSA
\citep[e.g.,][]{1993ApJ...402..560J,1994APh.....2..215B,1995ApJ...447..944K},
only particles in the exponential tail of the Maxwellian thermal
distribution will be able to cross the shock upstream to undergo
acceleration. The threshold momentum is:
\begin{equation}
  \label{eq:qinj}
  \pinj = \xinj p_\rmn{th} = \xinj \sqrt{\frac{2 \,k T_2}{m c^2}}.
\end{equation}
where typically $x_{\rm inj} \approx 3.5-4$. We adopt a fit to Monte
Carlo simulations of the thermal leakage process \citep{kang11}:
\begin{equation}
  \label{eq:xinj}
  x_\rmn{inj} \approx 1.17 \frac{u_2}{p_{\rm th}\,c} \left(1+
  \frac{1.07}{\epsilon_B}\right) \left(\frac{\mach}{3}\right)^{0.1}\,.
\end{equation}
where the Mach number of the shock ($\mach$) is the ratio of the
upstream velocity and sound speed, $\eb = B_0/B_{\perp}$, $B_0$ is the
amplitude of the downstream MHD wave turbulence, and $B_{\perp}$ is
the magnetic field along the shock normal. The physical range of $\eb$
is quite uncertain due to complex plasma interactions, although both
plasma hybrid simulations and theory suggest that $0.25 \la \eb \la
0.35$ \citep{1998AdSpR..21..551M}. In this paper, we adopt $\eb =
0.23$, which corresponds to a conservative maximum energy acceleration
efficiency for electrons (equation~\ref{eqn:energy_frac}) of $\sim 1$
per cent. We show in Appendix~\ref{sect:tests} how our results vary
with this parameter.

The thermal post-shock distribution is a Maxwellian: 
\begin{equation}
  \label{eq:MBdistibution}
  f_\rmn{th}(p) = 4\upi\, n_\rmn{th}\,
  \left(\frac{m c^2}{2 \upi\, k T_2}\right)^{3/2}\! p^2
  \exp\left(-\frac{m c^2\,p^2 }{2 \,k T_2}\right),
\end{equation}
where the gas number density ($n_\rmn{th}$) and temperature ($T_{2}$)
are derived from the shock jump conditions. In the test particle
regime, the CR power-law attaches smoothly onto the thermal post-shock
distribution, at $x_{\rm inj}$, which is the only free parameter:
\begin{equation}
  \label{eq:finj_lin}
  f_\rmn{CR,lin}(p) = 
  f_\rmn{th}(\pinj) \left(\frac{p}{\pinj}\right)^{-\alpi} 
  H(p-\pinj).
\end{equation}
Fixing the normalization of the injected CR spectrum by this
continuity condition automatically determines the normalization
constant $C_\rmn{inj}$. The slope of the injected CR spectrum is:
\begin{equation}
  \label{eq:ainj}
  \alpi = \frac{\rtot + 2}{\rtot - 1} = 
  \frac{(\gamad+1)\mach^2}{(\gamad-1)\mach^2+2}\,, 
\end{equation}
where $\gamad=5/3$ is the adiabatic index and
\begin{equation}
  \label{eq:rtot}
  \rtot = \frac{\rho_2}{\rho_0} = \frac{u_0}{u_2}
\end{equation}
denotes the shock compression ratio \citep{1978MNRAS.182..147B,
  1978MNRAS.182..443B, 1983SSRv...36...57D}. We assume here that the
upstream Alfv{\'e}n Mach number ${\cal M}_{\rm A} = u_{0}/v_{\rm A}
\gg 1$, so magnetic fields are dynamically unimportant. The number
density of injected CR particles is given by
\begin{equation}
  \label{eq:ninj}
  \Delta n_\rmn{CR,lin} = \int_0^\infty \!\!\!\!\! \dd p\, f_\rmn{CR,lin}(p)
  = f_\rmn{th}(\pinj)\, \frac{\pinj}{\alpi-1}.
\end{equation}
This enables us to infer the particle injection efficiency, which is
the fraction of downstream thermal gas particles which experience
diffusive shock acceleration,
\begin{equation}
  \label{eq:eta}
  \eta_\rmn{CR,lin}\equiv\frac{\Delta n_\rmn{CR,lin}}{\nth} = 
  \frac{4}{\sqrt{\upi}}\,\frac{x_\rmn{inj}^3}{\alpi-1}\,
  \rmn{e}^{-x_\rmn{inj}^2}.
\end{equation}
The particle injection efficiency is a strong function of $\xinj$ that
depends on both $\mach$ and $\eb$ (for instance, it changes by more
than an order of magnitude for $\eb=0.25-0.30$ at $\mach = 3$. We
discuss this further in Appendix~\ref{sect:tests}). The energy density
of CRs that are injected and accelerated at the shock (neglecting the
CR back reaction on the shock) is given by
\begin{equation}
\label{eq:CR_energy} 
  \Delta\eps_\rmn{CR,lin} =
  \eta_\rmn{CR,lin}(\mach)\,T_\CR(\mach,\pinj)\,\nth(T_2),
\end{equation}
and the CR energy injection and acceleration efficiency is:
\begin{equation}
  \zeta_\rmn{lin} =
  \frac{\Delta\eps_\rmn{CR,lin}}{\Delta\eps_\rmn{diss}},
   \quad\mbox{where}\quad
  \Delta\eps_\rmn{diss} = \eps_\rmn{th2} - \eps_\rmn{th0}\,\rtot^{\gamad}\,.
\label{eqn:energy_frac}  
\end{equation}
The dissipated energy density in the downstream regime,
$\Delta\eps_\rmn{diss}$, is given by the difference of the thermal
energy densities in the pre- and post-shock regimes, corrected for the
adiabatic energy increase due to gas compression.

\subsubsection{Models for the CR Spectrum} 
\label{sec:models} 
\begin{figure}
\begin{center}
{\includegraphics[width=0.99\columnwidth]{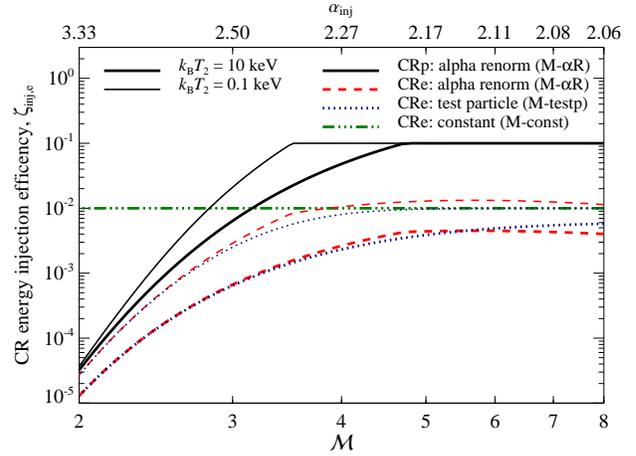}}
\end{center}
\caption{Acceleration efficiency as a function of shock strength for
  the spectral index renormalization model (`M-$\alpha$R', see
  \S\ref{sec:models} for details). Those are compared to the electron
  acceleration efficiency in various models: `M-$\alpha$R' (red
  dashed), the test particle acceleration model (`M-testp'), and a
  model with constant acceleration efficiency that is widely used in
  the literature (`M-const'). The thick lines represent shocks in a
  hot medium with a post-shock temperature $kT_2 = 10$~keV, and the
  thin lines represent shocks in colder media with $kT_2 = 0.1$~keV.}
\label{fig:etamax} 
\end{figure}

The above formalism for diffusive shock acceleration describes the
standard test particle scenario in which the relativistic particles
have no influence on the shock structure. However, these equations are
no longer valid at high Mach numbers, when they predict that the
fraction of energy which goes into relativistic particles (as in
equation~\ref{eqn:energy_frac}) can reach or exceed $100$ per cent. By
contrast, numerical studies of shock acceleration of ions suggest that
$\zeta_\rmn{lin}$ has a upper energy injection efficiency limit set by
$\zeta_\rmn{max} \simeq 0.1$ \citep{2013ApJ...764...95K}. This limit
arises from nonlinear effects, due to the back-reaction of the
accelerated particles upon the shock
\citep{1979ApJ...229..419E,1981ApJ...248..344D,1982A&A...111..317A,2000ApJ...533L.171M,2002APh....16..429B,2005MNRAS.361..907B,2005ApJ...620...44K}.

While non-linear models of particle acceleration certainly exist, for
our purposes they introduce needless complication: as we shall see,
due to the strong cooling processes at play, we are not strongly
sensitive to detailed features of the injected spectrum. We therefore
employ simple modifications of the test-particle picture to ensure
energy conservation is not violated. We require that the energy
injection efficiency $\zeta_{\rm inj}$ obeys an upper bound
$\zeta_{\rm max}$:
\begin{equation}
\label{eq:saturation}
\zeta_\rmn{inj} = \left[1 - \exp\left(-\frac{\zeta_\rmn{lin}}
  {\zeta_\rmn{max}}\right)\right]\,\zeta_\rmn{max}.
\end{equation}
where we set $\zeta_{\rm max} = 0.1$ for the CRp population, and
$\zeta_{\rm max}=0.01$ for the CRe population (equivalently, we can
set $\zeta_{\rm inj} \approx {\rm min}(\zeta_{\rm lin},\zeta_{\rm
  max})$) \footnote{Note that the subscript 'lin' denotes the injected
  and accelerated population in the test particle approximation, while
  'inj' refers to the injected and accelerated population after
  accounting for CR back reaction effects on the shock structure that
  leads to saturation of the accelerated CR energy density.}. If the
injected one-dimensional CRe distribution function at each time $t$
is:
\begin{equation}
  f_\rmn{inj,CRe}(p,t) = \Delta C_\e\,p^{-\alpi} H(p-p_{\rm cut})\,, 
\end{equation}
then there are 3 effective time-dependent variables $\Delta C_\e,
\alpi, p_{\rm cut}$ to satisfy the twin constraints of injected number
density (equation \ref{eq:eta}) and energy density
(equation~\ref{eq:CR_energy} and \ref{eq:saturation}). We consider two
simple models. Model `M-$\alpha$R' is motivated by models of
non-linear shock acceleration, and varies $(\Delta
C_\e,\alpha_\rmn{inj})$ (keeping $p_{\rm cut}=p_\rmn{inj}$ fixed),
while Model `M-testp' varies $(\Delta C_\e, p_{\rm cut})$, keeping the
test particle slope $\alpha_\rmn{inj}$ constant. We also account for
the different acceleration efficiencies of protons and electrons (due
to their different masses, which give rise to different gyroradii in
the non-relativistic regime) by assuming the CRe distribution function
mimics the CRp distribution function, but with a lower
normalization. The prescriptions for these two models is:

\begin{itemize} 
{\item {\bf Spectral index renormalization (`M-$\alpha$R')} We limit
  the acceleration efficiency by steepening the spectral index of the
  injected population $\alpi$ to $\alps$. The slope $\alpi$ affects
  $\zeta_{\rm inj}$ via the mean energy per particle,
  equation~\eqref{eq:Tcr}. This is motivated by models of non-linear
  shock acceleration where a subshock with a lower compression ratio
  (and hence steeper spectral index) forms
  \citep[e.g.,][]{2000ApJ...540..292E}. Given the assumed $\epsilon_B$,
  we find that for strong shocks where $\alpha \lesssim 2.3$ the
  spectral slope is steepened by a maximum of $\sim 10$ per cent in
  low temperature regimes ($kT\sim 0.1$~keV), while the steepening is
  much smaller for high temperature regimes ($kT\sim 10$~keV) that are
  more relevant for clusters. Since $p_{\rm inj}$ remains fixed, so
  does $n_{\rm CR}$, and we solve for the normalization constant
  $\Delta C_{\rm e}$ from equations~\eqref{eq:ncr} and
  \eqref{eq:ninj}:
\begin{equation}
  \label{eq:delta_Cp}
  \Delta C_\pro=\eta_\rmn{CR,lin}\,(\alps-1)\,\pinj^{\alps-1}\,.
\end{equation}
We then relate the injected CRes to the CRps by assuming:
\begin{equation}
  \label{eq:f_inj_relnorm}
  f_\e(P) = K_\rmn{e/p}(\alpha)\, f_\pro (P)\,,
\end{equation}
where $f_\e(P)$ and $f_\pro(P)$ are the CR distribution functions at
physical momentum $P$ for the electrons and protons, respectively. The
ratio between the electrons and protons can be fixed by requiring
equal CRe and CRp number densities above a fixed injection energy
$E_{\rm inj}$, which results in $ K_\rmn{e/p}(\alpha) =
X^{(1-\alpha)/2}$, where $X = m_{\rm p}/m_{\rm e}$ is the proton to
electron mass ratio \citep{schlickeiser02}. For the value of $\alpha
\sim 2.3$ (consistent with the injection spectral index at galactic
supernova remnants as traced by multi-frequency observations), this
yields $K_\rmn{e/p} \sim 0.01$, which is what is observed
locally. Note that the appropriate choice of normalization can be
highly uncertain --- for instance, supernova remnants give values of
$K_\rmn{e/p} \sim 10^{-4}$ which are significantly smaller
\citep{morlino09}. However, it does not significantly affect our
results in this paper, which hinge on the scaling of acceleration
efficiency with Mach number.}

{\item {\bf Modified test particle (`M-testp')} This model preserves
  the test particle slope $\alpha_{\rm inj}$ and instead modifies the
  cutoff $p_{\rm cut}$ of the power-law distribution, which is no
  longer necessarily equal to $p_{\rm inj}$
  \citep{2007A&A...473...41E} in the regime where the injection
  efficiency saturates. This was argued to mimic the effect of rapid
  Coulomb cooling in the non-relativistic regime
  \citep{2007A&A...473...41E}, though the physical motivation for this
  is less clear (our calculations explicitly track Coulomb
  cooling). Nonetheless, we include this model to allow comparison
  with previous published results. We then solve the two equations
  $\Delta \eps_\CRe(\alpi,q_{\rm cut}) = \zeta_{\rm inj,e} \Delta
  \eps_\rmn{diss}$ and $\Delta n_\CRe(\alpi,q_{\rm cut}) =
  \Delta\,\eta_\rmn{CR,lin}\,\nth(T_2)$ for $C_\e$ and $q_{\rm cut}$.}
\end{itemize} 

We regard the `M-$\alpha$R' model as our default model, since the
steepening of the spectral slope is physically well-motivated. In
practice, we shall see that these two models give almost identical
results, as the differences between them in the momentum regime
important for fossil reacceleration is negligible. We also adopt a
straw-man model where a constant energy acceleration efficiency
independent of Mach number is assumed:
\begin{itemize}
{\item {\bf Constant acceleration efficiency (`M-const')}} Similar to
the `M-testp' model, we solve for both $p_{\rm cut}$ and $\Delta
C_{e}$ with $\alpha$ fixed to the test particle slope for a
combination of $\epsilon_{e}, n_{\rm CRe}$; however, the energy
injection efficiency $\zeta_{\rm inj,e}=1$ per cent is assumed to be
constant and independent of Mach number.
\end{itemize} 
Unlike the previous two models, we do {\it not} regard this model as
physically realistic. It ignores what we know about the strong
dependence of acceleration efficiency on Mach number, and consequently
vastly overestimates the number of weak shocks visible if only direct
injection operates (see \S\ref{sec:flux}). Nonetheless, because it has
been widely adopted in the literature, we adopt it as a straw man
model to compare against previous results.

We show the acceleration efficiency for all 3 models in
Fig.~\ref{fig:etamax}. The `M-$\alpha$R' and `M-testp' models give
very similar results. They are comparable to the `M-const' model for
high $\mach$, but have much lower acceleration efficiencies at low
$\mach$.

\subsection{CR Electron Cooling}
The CRes cool through synchrotron and inverse Compton (IC)
emission\footnote{Hereafter, the initials 'IC' should be understood to
  denote both processes. For the purpose of calculating the fossil
  electron distribution function, we ignore synchrotron cooling. The
  CMB has an energy density equivalent magnetic field strength of
  $B_{\rm CMB} = 3.24 (1+z)^2 \mu{\rm G}$, and is generally dominant.}
and Coulomb interactions on timescales that are relatively short
compared to the dynamical timescale of a cluster. The finite time
resolution of our simulations imply that we inject electrons at
discrete time intervals, rather than continuously. We therefore
incorporate cooling in our simulations by considering the evolution of
a power-law of spectrum electrons instantaneously injected at time
$t_i$ and evolved forward to a later time $t_f$ (for further details,
see \citet{1999ApJ...520..529S}, and Appendix~\ref{app:cool}). If
there is no further injection of CRes, then the number of particles is
conserved:
\begin{equation}
\label{eq:num_dens_cons}
 \int_{\pf}^{\infty} f_\rmn{inj,CRe}(p',t_f)\,\dd p' =
 \int_{\pin}^{\infty} f_\rmn{inj,CRe}(p',t_i)\,\dd p'\,.
\end{equation}
The distribution function is then given by 
\begin{eqnarray}
\label{eq:num_dens_cons2}
f_\rmn{inj,CRe}(\pf,t_f(t_i)) \!&=& \! f_\rmn{inj,CRe}(\pin,t_i)
\left.\frac{\partial\pin}{\partial\pf}\right|_{t_f}\,,\, \rmn{where} \\
 f_\rmn{inj,CRe}(\pin,t_i) \!&=& \!
 f_\rmn{inj,CRe}(\pf-\Delta \pIC-\Delta \pC,t_i).
\end{eqnarray}
Here $\Delta \pIC$ and $\Delta \pC$ represents the shift in momentum
from $\pin$ to $\pf$ of a CRe, due to inverse Compton and Coulomb
cooling, respectively. It is derived by integrating the loss function
$b(p,t)$, defined by
\begin{equation}
  \label{eq:d_cool}
  \frac{\dd E}{\dd t} = - b(E,t) = - \beta\,b(p,t) = \beta\,\frac{\dd
    p}{\dd t} \,,
\end{equation}
where $E$ denotes the particle energy in units of $m_\rmn{e} c^2$ and
$\dd E/\dd t$ represents the loss of energy for each particle.

The {\it Coulomb losses} for CRes are described by
\begin{eqnarray}
  \label{eq:b_C}
  b_\rmn{C}(p,t) &=& \frac{3\,\sigma_\rmn{T}\,n_\e\,c}{2\, \beta^2}
  \left[\ln\left(\frac{\mae c^2 \beta \sqrt{\gamma-1}}{\hbar\,\omega_\rmn{plasma}}\right)\right.\nonumber \\
    &-&\ln(2)\left(\frac{\beta^2}{2}+\frac{1}{\gamma}\right)+\frac{1}{2}+
    \left.\left(\frac{\gamma-1}{4\gamma}\right)^2\right]\,, \\
\rmn{where} &\,& \beta = p/\sqrt{1+p^2}\,,\quad\rmn{and}\qquad \gamma=\sqrt{1+p^2}\,.\nonumber
\end{eqnarray}
Here $\omega_\rmn{plasma} = \sqrt{4\pi e^2 n_\e / \mae}$ is the plasma
frequency, and $n_\e$ is the number density of free electrons. The
details of the Coulomb cooling are given in
Appendix~\ref{app:cool}. An important feature to note is that $b
\propto p^{-2}$ for non-relativistic (NR) electrons, while $b
\approx$~const for relativistic (R) electrons, implying $t_{\rm cool}
\propto p^{3}$ (NR) and $t_{\rm cool} \propto p$ (R). We will see how
this shapes the fossil spectrum in \S\ref{sect:analytic}.

The {\it inverse Compton} losses are given by
\begin{equation}
  \label{eq:b_IC}
  b_\IC(p,z) = \frac{4}{3}\frac{\sigma_\rmn{T}}{\mec}\,\frac{p^2}{\beta}\,U_\CMB
  = b_{\IC,0}\,\frac{p^2}{\beta}\,(1+z)^4\,,
\end{equation}
where the energy density of the CMB, $U_\CMB =0.26
[\eV \cm^{-3}](1+z)^4$, and the Thomson cross section, $\sigma_\rmn{T}=
8\pi e^4/3(\mae c^2)^2$. Similarly to Coulomb cooling we derive the
momentum evolution of a particle subject to IC losses through
\begin{equation}
  \label{eq:b_IC_evolu}
\beta\,\frac{\dd p}{p^2} = -b_{\IC,0} (1+z)^4\dd t\,.
\end{equation}
When a time $\Delta t = (t_f-t_i)$ has elapsed, all energetic CRes
with a momentum $p \geq \pIC$ have cooled to a lower momentum $p <
\pIC$. We derive $\pIC$ by integrating equation~\eqref{eq:b_IC_evolu}
from the redshift $z_i=z(t_i)$, where the electrons are injected, to a
later time $z_f=z(t_f)$ where they are evaluated. The details of the
IC cooling are given in Appendix~\ref{app:cool}.

The total electron spectrum is derived from the sum of all
individually cooled injected spectra (denoted by summation index $j$)
\footnote{For each simulated cluster, the CRe spectrum is followed on
  500 randomly selected SPH particles. Convergence studies show only
  marginal differences with increasing number of SPH particles.},
starting from the time of injection $t_i$ until a later time $t_f$,
\begin{eqnarray}
  f_\CRe(\pf,t_f) = \sum_j f_\rmn{inj,CRe}(\pf,t_f,t_j)\,.
\end{eqnarray}

It is important to note that our injection time steps of $\Delta t$ =
100~Myr, 1~Gyr are in fact often longer than the cooling time of
electrons at both low and high energy extremes (see
Fig.~\ref{fig:timescales}). This low time resolution can have the
effect of severely overestimating the impact of cooling. Even so, it
turns out that our simulations generally give results accurate to
within a factor $\sim 2$ for the reaccelerated spectrum, compared to
calculations where injection and cooling are fully resolved. We
discuss this in detail in \S\ref{sect:time_resolution}.

\section{Fossil Electrons: Results and Physical Interpretation}
\label{sect:spectrum_results} 
We present our primary results for the relic electron distribution
function in this section. In \S\ref{sect:spectrum_results}, we analyze
a representative CR electron spectrum in the cluster outskirts,
choosing a radial range of $(0.8-1.0)\,\rvir$. Accounting for
projection effects, this appears to be consistent with the observed
distribution of projected distances of elongated relics, which ranges
from 0.5--3 Mpc, with a mean of 1.1 Mpc \citep{2012A&ARv..20...54F}.
In Sect.~\ref{app:radial}, we will show that our main results are
insensitive to the exact choice of this radial range. We focus on the
underlying physics which gives rise to the spectrum, and show in
\S\ref{sect:analytic} that a simple analytic model can fit our
simulation results in the momentum regime resolved by our simulations.

\subsection{Representative Fossil Spectrum: Simulation Results}
\label{sect:sim_spectrum} 
\begin{figure}
  \includegraphics[width=1.0\columnwidth]{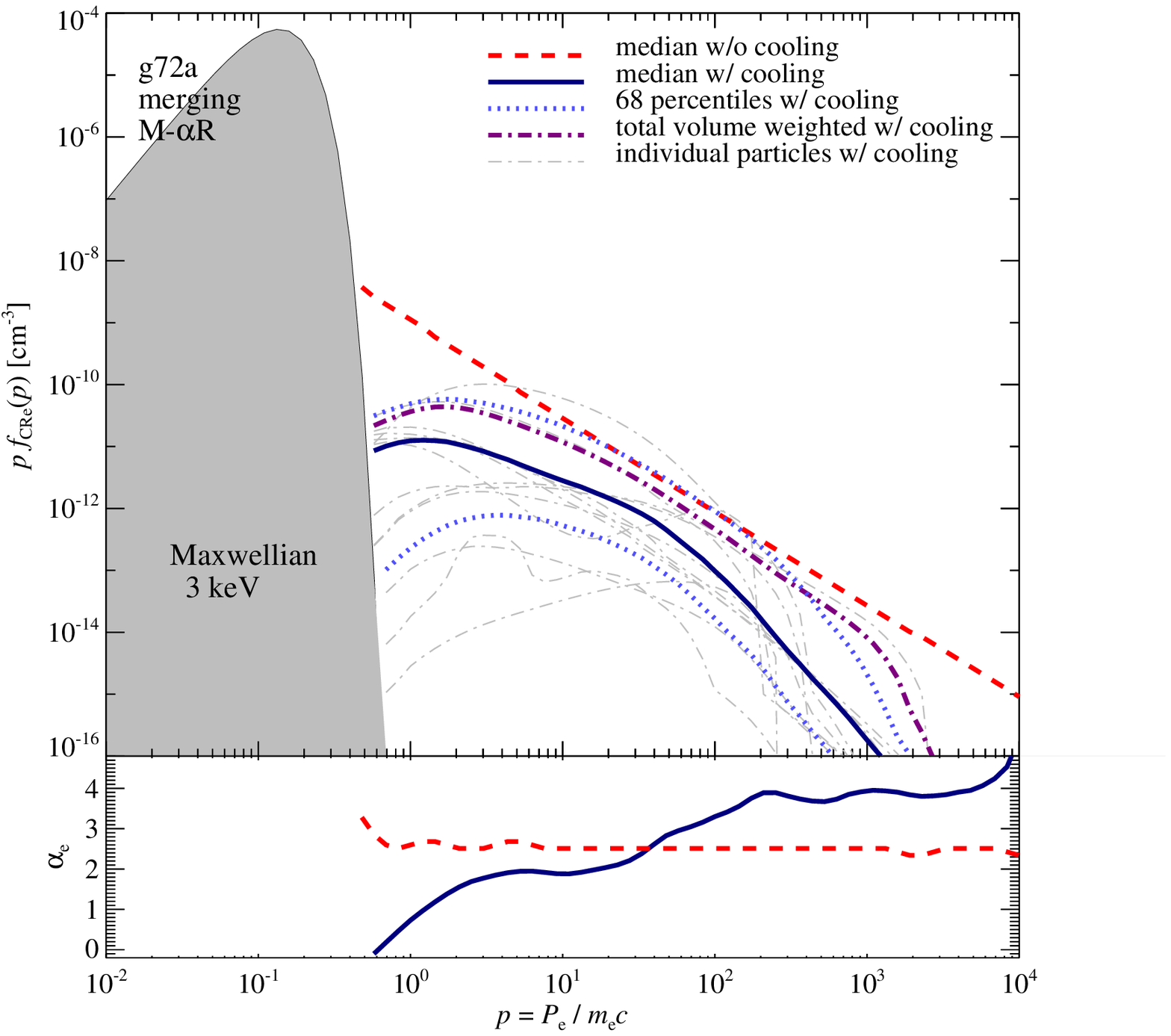}
  \caption{Representative electron momentum spectrum $f_\CRe$ of a
    post-merging cluster in the virial region between
    $(0.8-1.0)\,\rvir$. Top: we show the median injected spectrum (red
    dashed), the median cooled spectrum (dark blue solid) and its 68
    percentiles (light blue dotted). Overplotted are representative
    sample spectra of individual SPH particles (grey dash-dotted). The
    total volume-weighted spectrum (purple dash-dotted). The cosmic
    rays are compared to the thermal population described by a
    Maxwellian with a temperature of $kT_2=3\,\rmn{keV}$. Bottom:
    spectral index of the median electron spectrum.}
  \label{fig:e_spec_represent}
\end{figure}

Let us first consider the spectrum from a typical cluster, the
post-merger cluster g72a, generated using our `M-$\alpha$R' model
(from Table \ref{tab:cluster_sample}, note that 9 of the 14 simulated
clusters are in the process of merging or had a recent merger). We
focus on a particular case because the fossil spectra all have generic
features which can be understood from a physical standpoint. In
Fig.~\ref{fig:e_spec_represent}, we show results both from volume
weighting the individual spectra of SPH particles as well as the
median CRe spectrum. The normalization of the two spectra differ
significantly. The volume-weighted spectrum is dominated by about
$1/5$ of the particles which also make up a similar fraction of the
total volume in the cluster outskirts, except for high momenta ($p
\gtrsim10^3$) where the particle fraction that dominates the volume
weighted spectrum becomes progressively smaller. Here, we explore
cluster outskirts in SPH simulations with the aim of providing a
converged and robust result for the fossil electron spectrum. We
decided in favor of a conservative approach that takes our numerical
limitations into account and focus on the median spectrum, which is a
robust statistic relatively insensitive to the tails of the
distribution. We defer a detailed study of the full distribution
function to future work.

Note that the cooled spectrum deviates significantly from
the injected spectrum (shown with a red dashed line). The cooled
spectrum can be characterized by four regimes: (1) the
sub-relativistic Coulomb cooling regime ($p \lesssim 1$) where the CRs
injected in the most recent shocks dominate the population; (2) the
relativistic Coulomb cooling regime, which includes a peak at $p \sim
1$) due to the transition from non-relativistic to relativistic
cooling; (3) the adiabatic regime ($10 \lesssim p \lesssim 10^2$)
where the cooling time is long ($\sim$ few Gyrs); (4) the radiative
cooling regime ($p \gtrsim 10^2$) where the synchrotron and inverse
Compton losses start to steepen the CRe spectrum.
 
In the lower panel of Fig.~\ref{fig:e_spec_represent} we show the
spectral index of the CRe spectrum, which is the same as that for the
CRp. The {\it injected spectra} for CRp's has a concave shape with a
spectral index $\sim 2.7$ at $p\sim 1$ which has flattened to $\sim
2.4$ at $p\sim 10^{3}$. This spectral shape is a consequence of the
cosmological Mach number distribution that is mapped onto the CR
spectrum \citep{2006MNRAS.367..113P}. It can be understood
qualitatively as follows. The characteristic Mach number declines with
time in a $\Lambda$CDM universe, due to the slow-down in structure
formation\footnote{By contrast, in an Einstein-de Sitter
  $\Omega_{m}=1$ universe, structure formation is self-similar
  \citep{bertschinger85}, and the cosmic Mach number should not show
  any evolution, aside from non-gravitational events such as
  reionization and effects from finite mass resolution in the
  simulations \citep{2006MNRAS.367..113P}.}. Thus, shocks at early
times have higher Mach numbers and harder spectra; late time shocks
have softer spectra. However, early shocks also have lower
normalizations, because they take place in a colder
medium. Specifically, $p_{\rm inj} \propto T^{1/2}$---the power-law
attaches to the thermal distribution at lower momenta in a colder
medium. Thus, the normalization $C \propto p_{\rm inj}^{\alpha-1}
\propto (x_{\rm inj} T^{1/2})^{\alpha -1}$ is lower; for gas with
$T_{1}=10^{4}$K and $T_{2}=10^{7}$K, $C_{2}/C_{1} \approx
(10^{3})^{0.75} (x_{\rm inj,2}/x_{\rm inj,1})^{1.5} \approx 100$,
adopting $\alpha=2.5$, $x_{\rm inj,2}\sim x_{\rm inj,1}$, and assuming
a constant number density of thermal electrons for this
order-of-magnitude argument. We shall soon see (\S\ref{sect:analytic}
and Fig.~\ref{fig:density_evol}) that the competing effects of
structure formation and the expansion of the universe conspire to keep
gas density that ends up in the cluster outskirts relatively constant,
so adiabatic evolution does not significantly alter this
conclusion. The upshot is that late-time weak shocks dominate at low $p$
due to their higher normalization (hence, the spectral slope at low $p$
is softer), while early-time strong shocks dominate at high $p$ due to
their harder spectra (hence, the slope here is harder). What is
important to note is that the spectral index in the adiabatic regime
$p\sim 10-100$ is relatively soft ($\alpha \approx 2.5$), implying
that primarily weak shocks $\mach \sim 3$ contribute.

\begin{figure}
  \includegraphics[width=0.99\columnwidth]{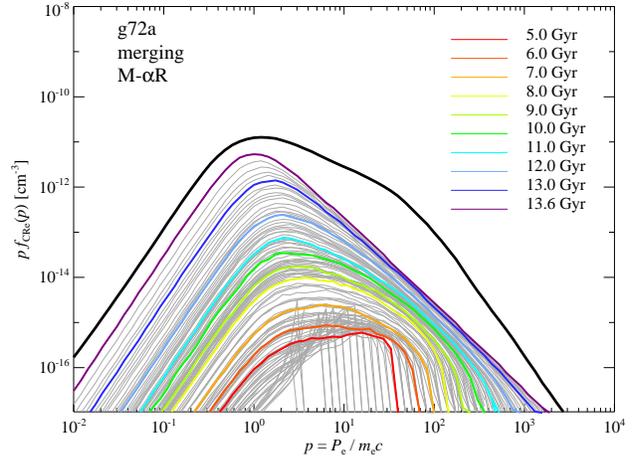}
  \caption{Differential spectral build up of the median electron
    distribution function, $f_\CRe$, as a function of cosmic time, for
    a post-merging cluster in the region between $(0.8-1.0)\,\rvir$.
    The black solid line shows the total $f_\CRe$. The thin grey lines
    show the individual differential contributions to $f_\CRe$ from
    different times, where we highlight the contributions spaced by
    Gyr intervals with different colors. The spectral features are
    shaped by different cooling processes: non-relativistic Coulomb
    cooling for $p<1$, a broad feature indicating the transition from
    relativistic Coulomb cooling to the adiabatic regime for
    $1<p\lesssim10^2$, and inverse Compton cooling at higher momenta.}
    \label{fig:e_specZ}
\end{figure}

It is also instructive to consider the relative contribution of
different cosmological epochs to the spectra, shown in
Fig.~\ref{fig:e_specZ}. There are two important effects: the
decreasing normalization of spectra injected at early times (for the
reasons discussed above), and the strong effects of cooling. Both
conspire to decrease the importance of early time injection. Thus, the
first effect means that most electrons in the peak and 'adiabatic'
regime $1 \lsim p \lsim 100$ came from the last $\sim 3$~Gyr. In the
regime $p \lsim 1$, where sub-relativistic Coulomb cooling dominates,
the contribution is even more recent--essentially within the last
cooling time, $\sim {\cal O}(10^{8})$~yrs. We now turn to
understanding the shape of this spectrum in detail.

\subsection{Analytic Model for Fossil Electrons}
\label{sect:analytic} 
We develop an analytic model for the fossil distribution function
$f_\CRe(p)$ in the presence of multiple cooling processes. There
are two clear limiting cases: an ``impulsive injection'' scenario where
an initial population of relativistic particles cools passively, and a
``steady injection'' scenario where a steady state balance between
injection and cooling is achieved. The ``impulsive'' scenario is
appropriate when the timescale between shocks is greater than the
cooling time. The finite time resolution of our post-processed
simulations means that our simulation results are a linear
superposition of such impulsive solutions. The ``steady state'' scenario
is appropriate when the injection rate of relativistic particles is
fairly constant, and $t_{\rm cool}(p) \ll t_{\rm H}$ (where
$t_{\rm H}$ is the Hubble time) so that the population equilibrates on
a short timescale.

\begin{figure*}
  \includegraphics[width=\columnwidth]{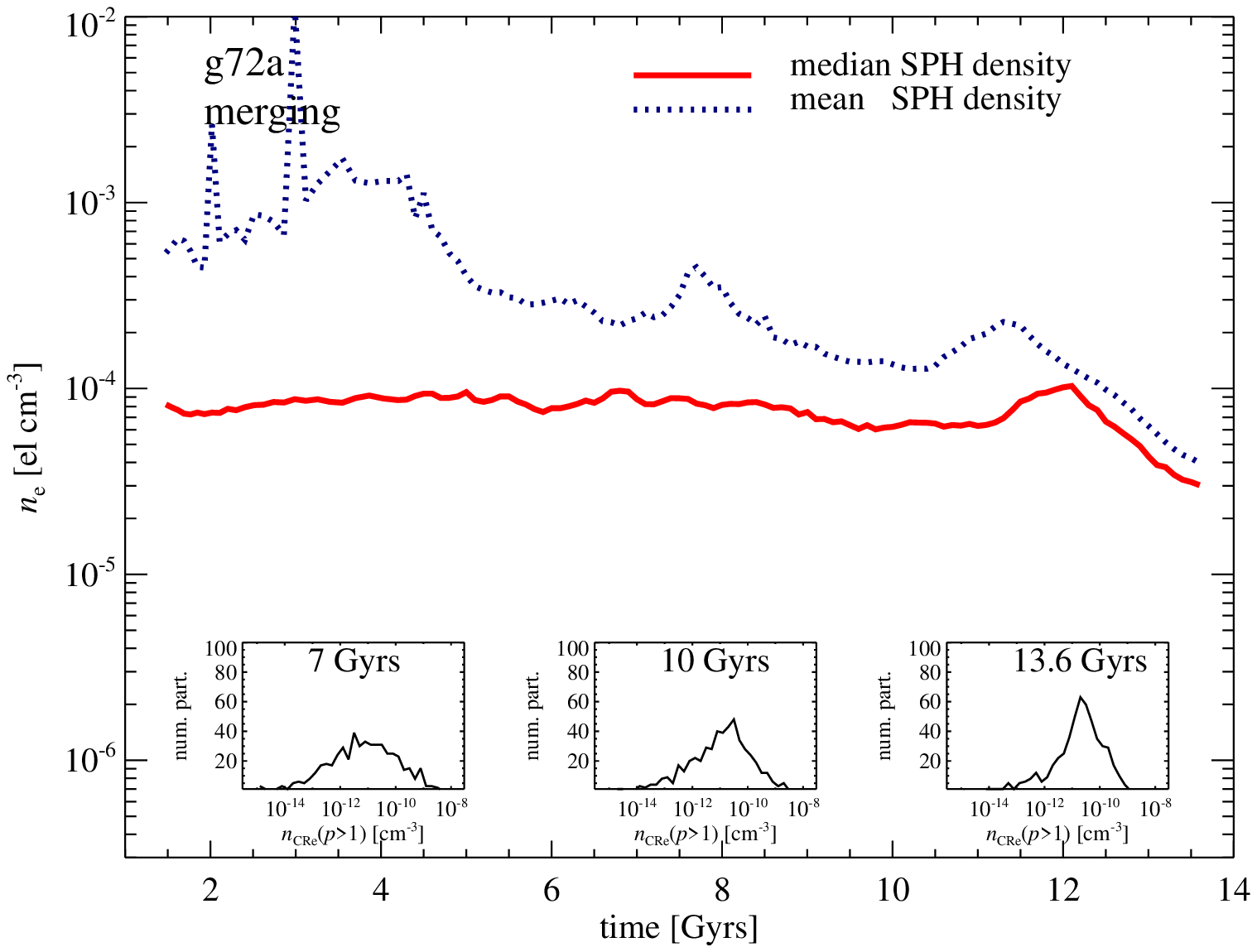}
  \includegraphics[width=\columnwidth]{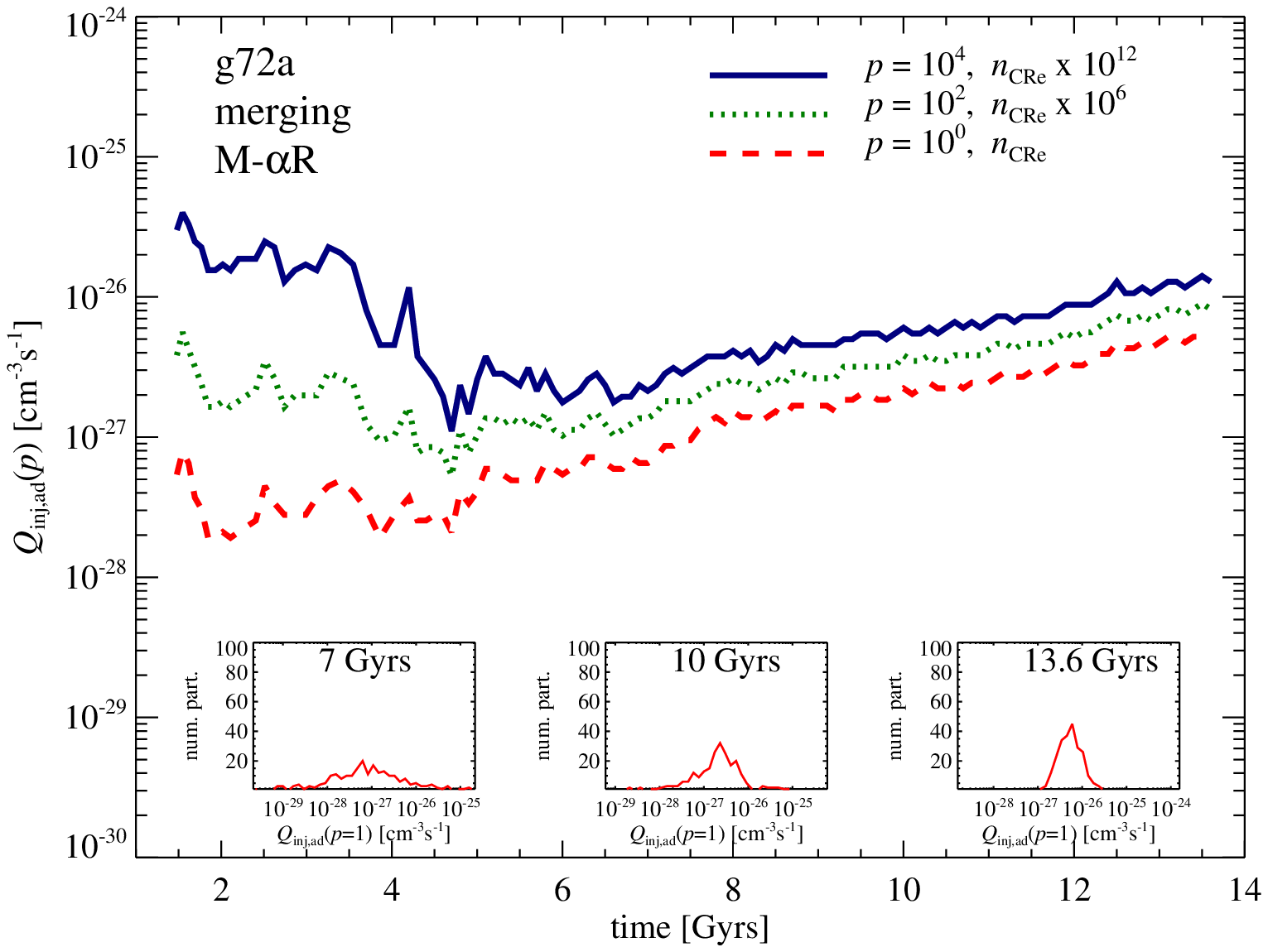}
  \caption{Lagrangian evolution of the thermal electron number
    density, $n_\rmn{e}$ (left), and the injection rate density of CR
    electrons, $Q_{\rm inj,ad}$ (right). We consider a representative
    sample of SPH particles that end up at the virial region of a
    massive, post-merging cluster (g72a) between $(0.8-1.0)\,\rvir$ at
    $z=0$ and follow their evolution as a function of time since the
    Big Bang. Left: the mean (blue dotted) and median (red solid)
    $n_\rmn{e}$. Their difference is due to its non-Gaussian
    distribution caused by clumping, which biases the mean
    upwards. Right: we show $Q_{\rm inj,ad}$ for three different
    momenta: $p=1$ (red dashed ), $p=100$ (green dotted, weighted by
    factor $10^6$), and $p=10^4$ (blue solid, weighted by factor
    $10^{12}$). The three small insets in each panel show CR electrons
    with a momentum $p>1$ at different times; the density distribution
    of the fossil CR electrons (left), and the injection rate density
    (right). The variance is caused by different shock strengths,
    injection efficiencies, thermal densities and decreases with
    time.}
\noindent \hrulefill
\label{fig:density_evol}
\end{figure*}

Naively, since the CRs are injected at discrete shocks, one might
expect the impulsive approximation to be most appropriate. In fact,
the opposite is true. The timescale on which the shock injection
process changes is the dynamical time, $t_{\rm inj} \sim t_{\rm dyn}
\sim t_{\rm H}/\sqrt{\Delta} \sim 0.6 \, \Delta_{300}^{-1/2}$ Gyr,
where $t_{\rm H}=10$~Gyr, $\Delta_{300}=\Delta/300$ and
$\Delta=\rho/\bar{\rho}$. On, the other hand, as we shall soon see,
the main regime of interest where cooling significantly modifies the
spectrum is the trans- and sub-relativistic regime of Coulomb cooling,
where $t_{\rm cool} \sim 0.6 \, p^{3} (n_\e/ 3 \times 10^{-5} \, {\rm
  cm^{-3}})^{-1}$ Gyr for $p < 1$. The fact that $t_{\rm cool} \lsim
t_{\rm inj}$ implies that the injection rate is roughly constant on
the cooling timescale over which the population equilibrates.
Moreover, the Coulomb cooling rate itself---which depends on the gas
density---changes on the dynamical time $t_{\rm
  dyn}$. Figure~\ref{fig:density_evol} shows the Lagrangian evolution
of the physical gas density that ends up in the virial region of a
cluster. The presence of clumping implies a non-Gaussian density
distribution that biases the mean of the distribution upwards in
comparison to the median, the latter of which also shows a much
smoother density evolution. In fact, the physical density does not
even evolve significantly on timescales $t\gg t_{\rm
  dyn}$\footnote{For want of a better explanation, we view this as
  coincidental cancellation between the competing effects of Hubble
  expansion and structure formation.}. Thus, on the short
equilibration timescale $t_{\rm cool}$ for the sub and
trans-relativistic regimes, the injection and cooling rates are
roughly constant, and a steady state solution is valid. However,
before constructing these solutions, it is worthwhile to examine the
injection and cooling processes in more detail.

The right-hand panel of Fig.~\ref{fig:density_evol} shows the
Lagrangian evolution of the median CRe injection rate, $Q_{inj,ad}\sim
\dot{n}_{CRe}$, of gas ending up in the outskirts of cluster g72a,
which experienced a merger about 2 Gyrs ago. The merger is clearly
visible in the density but not in the injection rate. This is because
(1) only a fraction of the SPH particles that we trace in the
simulations experience shocks induced by the merger, with a minor
impact on the median injection rate, and (2) we account for the
adiabatic cooling of the injected particles, which suppresses the
injection rate for the merger with a factor $\sim C_\inj/C_f
(\rho_f/\rho_\inj)^{((\ainj+2)/3)}\sim 4\times4^{-4.5/3}\sim0.5$. We
can also see that the injection rate is fairly steady over
cosmological timescales. Similarly, we have also found that the
injection rate of a cool core cluster that has not experienced a major
merger in the last 7~Gyrs is either decreasing slowly or constant with
time. Note that this is {\it only} true for the median spectrum (or
any other spectrum which is averaged over a large number of fluid
elements). An individual fluid element can have a more stochastic
injection history, and the fossil electron population in the outskirts
of the cluster can vary spatially. This implies that when a weak shock
propagates across a cluster, not all regions will light up with equal
intensity. This spread in the CRe population is shown in the insets on
the left panel of Fig.~\ref{fig:density_evol}, while the spread in the
injection rate is shown in the right panel. These distributions narrow
with time, but can still span 1-2 dex at $z=0$.

For now, we focus on understanding the median
spectrum. Figure~\ref{fig:density_evol} shows that the median density
over the last $\sim$ Gyr is $n_\e \sim 3 \times 10^{-5} \, {\rm
  cm^{-3}}$ (corresponding to $\Delta \sim 300$), while the median CRe
injection rate density is $\dot{n}_{\rm CRe}(p > 1) =
\int_{1}^{\infty} Q_\rmn{inj,ad}(p) \,\dd p \sim Q_\rmn{inj,ad}(p=1)
\sim 4\times10^{-27} {\rm cm^{-3} \, s^{-1}}$, where
$Q_\rmn{inj,ad}(p)=f_\rmn{inj,CRe}(p,t)/\Delta t\,
(\rho_f/\rho_\inj)^{(\alpha+2)/3}$. We can understand this from simple
order of magnitude arguments. From Fig.~\ref{fig:e_spec_represent},
the spectral index for the adiabatic spectrum is $\alpha \sim 2.5$,
corresponding to $\mach \sim 3$ shocks. This is consistent with Fig.~1
of \citet{kang11}, where they find that the kinetic energy flux of
shocks has a sharp drop-off for shocks with $\mach \lsim 3$. A $\mach
\sim 3$ shock injects CRs at the cluster periphery with number density
$n_\CRe(p > 1) \sim n_\e \eta_{\rm CR,lin} (p_{*}/\pinj)^{-\alpha+1}
\sim 3\times10^{-11} \, {\rm cm^{-3}}$, where $\eta_{\rm CR,lin} =
n_{\rm CR}/n_\e \sim 4\times10^{-7}$ is the injection efficiency for a
$\mach=3$ shock (where $\pinj \sim 4.4\, p_{\rm therm}$), and the
factor of $(p_{*}/\pinj)^{-\alpha+1}$ converts $n(p > \pinj)$ to
$n(p>p_{*})$, and $p_{*} \sim 1$. Since the shocks operate on a
dynamical timescale $t_{\rm dyn} \sim t_{\rm H}/\sqrt{\Delta}\sim
0.6\,\rmn{Gyr}\,\Delta_{300}^{-1/2}$, the estimated injection rate is
$\dot{n}_{\rm CR}(p > 1) \sim {n}_{\rm CRe}(p > 1)/t_{\rm dyn} \sim
2\times10^{-27} \, {\rm cm^{-3} \, s^{-1}}$, which agrees with the
simulation results within a factor of two.

As for cooling, Fig.~\ref{fig:timescales} illustrates the cooling time
as a function of energy, for $n_\e \sim 10^{-5} {\rm cm^{-3}}$ at
$z=0$. There is a quasi-adiabatic regime $p_{1} < p < p_{2}$ where
$p_{1} \sim 10$, $p_{2} \sim 10^2$ where $t_{\rm cool} > t_{\rm H}$;
otherwise, for $p < p_{1}$, Coulomb cooling dominates, while for $p >
p_{2}$, inverse Compton cooling dominates. For our purposes, Coulomb
cooling is the most important cooling process. Inverse Compton cooling
merely shifts particles from the high energy tail to the adiabatic
regime (which acts as a 'road-block'), where they are still available
for reacceleration. Moreover, the relative number of affected
particles in the power-law tail is relatively small. By contrast,
Coulomb cooling affects the low-energy regime where most of the
particles are by number, and can cause them to migrate to low thermal
momenta, where they are no longer available for reacceleration. Note
that the Coulomb cooling function (equation~\ref{eq:b_C}) undergoes a
rapid transition from relativistic energies ($b_{\rm C} \approx {\rm
  const}$) to sub-relativistic energies ($b_{\rm C} \propto p^{-2}$),
which implies cooling times $t_{\rm cool} \propto p$ (for $1\lesssim p
\lesssim 30$, where the upper limit depends on density and redshift)
and $t_{\rm cool} \propto p^{3}$ (for $p < 1$). This rapid
thermalization of sub-relativistic particles creates a peak in the
steady state spectrum at $p \sim 1$.

\begin{figure}
  \includegraphics[width=1.0\columnwidth]{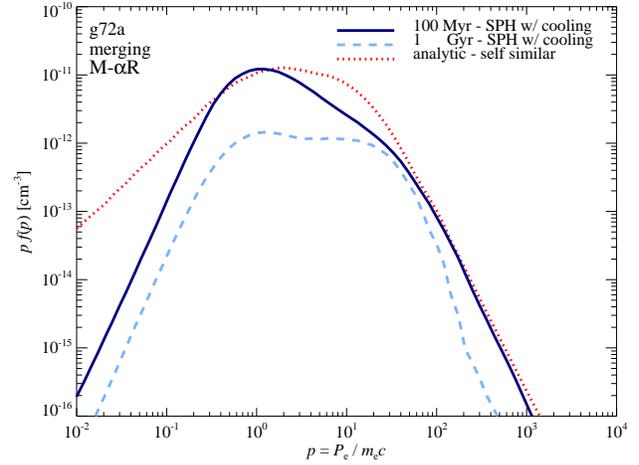}
  \caption{Comparison of the self-similar model (red dotted) of
    equation~\eqref{eq:f_self_similar} to the median CRe spectra,
    $f_\CRe$, of our simulated post-merging cluster g72a. We show the
    median $f_\CRe$ for two different time resolutions, 100~Myr (dark
    blue solid) and 1~Gyr (light blue dashed), indicating impressive
    convergence at $p\gtrsim100$, which is the relevant energy regime
    for radio relic emission.}
\label{fig:self_similar_g72a} 
\end{figure}

Let us now consider a toy model where the injection and Coulomb
cooling rates are constant. As previously mentioned, this is a
reasonable approximation, as they both change on a dynamical time
$t_{\rm dyn}$ which is longer than the equilibration time $t_{\rm
  cool}$. For the same reason, we can ignore adiabatic changes to the
electron spectrum. The magnetic fields which govern synchrotron
cooling also change on a dynamical time (for adiabatic changes, or a
turbulent dynamo), while inverse Compton cooling changes on
cosmological timescales. It is straightforward to solve for the steady
state solution, using the Vlasov equation. However, this is only
appropriate in the low and high energy regimes, where cooling (and
hence equilibration) timescales are short. In the adiabatic regime
($p_{1} < p < p_{2}$), the electron population is time-dependent---it
simply grows with time. Fortunately, when the injection and cooling
function are time-independent, one can construct self-similar
solutions which connect the steady and non-steady populations
\citep{1999ApJ...520..529S}. In these solutions, the overall
normalization varies with time, but the shape of the energy spectrum
remains the same if $p$ is scaled by some characteristic value. It is:
\begin{eqnarray}
  \label{eq:f_self_similar}
  f_{\rm self-sim}(p,t)&\approx&
  \frac{Q_{\rm inj,ad}(p)\,\Delta t}{{\ainj-1}}\,
  \left(\frac{\pCt}{p}+\frac{p}{\pICt}\right)^{-1}
  \nonumber\\ &\times&
  \left\{2-\left[\frac{1}{2}\left(1-\frac{p}{\pICt}+
    \left|1-\frac{p}{\pICt}\right|\right)\right]^{\ainj-1}\right.
  \nonumber\\ &-&
  \left.\left(1+\frac{\pCt}{p}\right)^{-(\ainj-1)}\right\}\,.
\end{eqnarray}
The approximation in equation~\eqref{eq:f_self_similar} is valid as
long as $\pICt\gg \pCt$, where $\pICt \sim 1/(b_\rmn{IC,0}\,\Delta t)$,
and $\pCt \sim b_\rmn{C}(p,t)\,\Delta t$
\footnote{We introduce an order unity factor to $\pICt$ and $\pCt$ to
  correct for the approximation that the loss and production terms are
  time-independent. For most of our simulated clusters we adopt a
  characteristic timescale for the cooling processes of $\Delta t\sim
  5-10$~Gyrs, $\pICt \sim 10$, and $\pCt \sim 1-5$.}. This expression
explicitly makes the approximation that the injection rate $Q_{\rm
  inj,ad}$ and cooling rates $b_{\rm C},b_{\rm IC}$ are independent of
time. It thus only requires two time-independent input parameters,
$Q_{\rm inj,ad},n_{\rm e}$. In Appendix~\ref{sect:CRinjection} we
provide a power-law model for $Q_{\rm inj,ad}$ with the spectral index
$\ainj$. In Fig.~\ref{fig:self_similar_g72a}, we compare the results
of our simulations of g72a (which have 100~Myr time resolution) to the
self-similar model in equation~\eqref{eq:f_self_similar}. The
agreement is remarkable, given that injection and cooling rates are
not exactly constant. We also show a calculation with lower
time-resolution $\Delta t=1$ Gyr, where the agreement is worse,
especially in the momentum regime where IC cooling is substantial. The
reason for this is that there are less SPH particles, which got
recently injected. Hence IC cooling modified the spectrum more
severely and removed all CRes with a momentum above $\pIC$. As a
result, median spectrum is biased low at high energies. We shall
return to this issue of finite time resolution in
\S\ref{sect:time_resolution}.

\section{Reacceleration} 
\label{sec:reaccel}    

\subsection{Which Momentum Regime Matters?} 
\label{sec:momentum_regime} 

\begin{figure}
  \includegraphics[width=1.0\columnwidth]{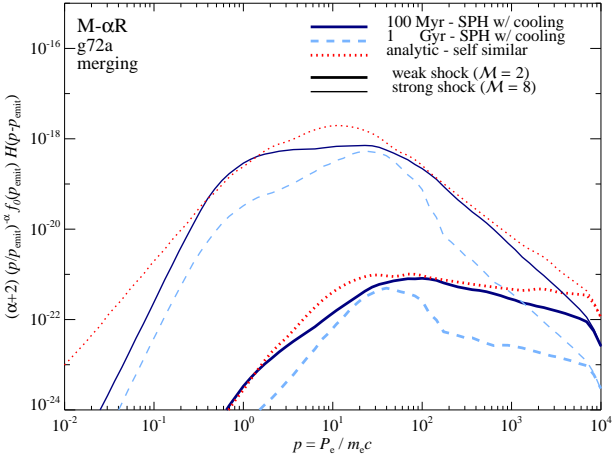}
  \caption{We show the contribution per logarithmic interval in
    momentum to the integral of the downstream distribution function
    of reaccelerated CR electrons in equation~\eqref{eqn:reaccel}. For
    a strong shock ($\mathcal{M}\gtrsim4$), there is a broad peak of
    fossil electrons with $1\lesssim p\lesssim100$ dominate the
    contribution to reaccelerated CR electron distribution, while for
    a weak shock, higher momenta with $p\gtrsim10$ contribute
    equally.}
\label{fig:p_contrib_reaccel}
\end{figure}

Downstream of a shock, the CR distribution function can be written as
\citep{1978MNRAS.182..443B,drury83}:
\begin{eqnarray}
f_{2}(p) &=& (\alpha+2)\,p^{-\alpha}\,\int_{p_{\rm inj}}^{p} 
p^{\prime \alpha-1} f_0(p^{\prime})\,\dd p^{\prime} \nonumber\\
&+& f_{\rm inj} \left( \frac{p}{p_{\rm inj}} \right)^{-\alpha} H(p-p_{\rm inj})\,,
\end{eqnarray}
where $\alpha$ is the test particle power-law slope and $H(p-p_{\rm
  inj})$ is the Heaviside step function, and $f_0(p^{\prime})$
represents the fossil CR electron population upstream of the merger
shock, mimicking the conditions for a radio relic. The first and
second terms on the right hand side refer to the contribution of relic
and freshly injected CRs respectively; we focus on understanding the
former in this section. This first term can be written in a more
physically instructive way as \citep{drury83}:
\begin{equation}
f_{2}^{\rm reaccel}(p) = \int_{p_{\rm inj}}^{\infty} (\alpha+2) 
\left( \frac{p}{p^{\prime}} \right)^{-\alpha} f_0(p^{\prime}) 
H(p-p^{\prime}) \dd({\rm log} p^{\prime})\,.
\label{eqn:reaccel}
\end{equation}
Thus, the downstream spectrum is a convolution of the upstream
spectrum with a truncated power-law. Note that since particles are
conserved and do not lose energy during the acceleration process, we
expect $n_{2}^{\rm reaccel}(p>p_{\rm inj}) = r n_0^{\rm
  fossil}(p>p_{\rm inj})$, where $r=\rho_{2}/\rho_{0}$ is the shock
compression ratio. Indeed one can derive equation~\eqref{eqn:reaccel}
from this assumption. These statements are reminiscent of the manner
in which a thermal Maxwellian tail with $p > p_{\rm inj}$ is converted
into a power-law---nothing about the acceleration process relies upon
a Maxwellian distribution, and the same conclusions hold for a
non-Maxwellian tail.

For concreteness, an illustrative example can be useful. Consider the
case when the relic spectrum is a power-law $f_{0}(p)=f_{0,
  *}(p/p_{\rm *})^{-\beta}H(p-p_*)$
\citep{kang11,kang12}. Substituting $f_{0}(p)$ into
equation~\eqref{eqn:reaccel} yields for $\alpha \neq \beta$
\citep{kang11}:
\begin{eqnarray}
\label{eqn:f-power-law}
f_{2}(p) &=& \frac{\alpha+2}{(\alpha-\beta)} 
\left[ 1 - \left(\frac{p}{p_{\rm *}} \right)^{-\alpha+\beta} \right] f_{0}(p) \\ 
&\approx& \frac{\alpha+2}{|\alpha-\beta|} f_{0, *} 
\left(\frac{p}{p_{\rm *}} \right)^{-\gamma}, \  \ \ p \gg p_{\rm *}\,, 
\label{eqn:f-power-law-approx}
\end{eqnarray}
where $\gamma={\rm min}(\alpha,\beta)$. Thus, for $p \gg p^{*}$, the
reaccelerated distribution function asymptotically becomes a
power-law, with a spectral slope corresponding to the shallower (i.e.,
harder) of the initial spectrum and the reaccelerating shock. Of
course, our fossil electron spectrum is very different from a
power-law, due the effects of cooling: it is a sharply peaked
function, with a continuously varying slope. Given that our
simulations are only accurate in a limited momentum range
$p^{\prime}$, we can ask: what range of $p^{\prime}$ must crucially be
resolved? Naively, one might assume that that since reacceleration
conserves number density, it suffices to resolve the peak of
$p^{\prime}f(p^{\prime})$, where most particles reside. However, the
additional weighting by a power-law in equation~\eqref{eqn:reaccel}
means that this condition is insufficient: although $p^{\prime}
f(p^{\prime})$ robustly peaks at $p^{\prime} \sim 1$, electrons here
may be 'too far from the action', once multiplied by the lever arm
$(p_{\rm emit}/p^{\prime})^{-\alpha}$.

Let us consider which initial momentum regime contributes most to
observed synchrotron emission of reaccelerated fossils. The
characteristic synchrotron frequency\footnote{Here we use the
  monochromatic approximation of synchrotron emission \citep[see
    App. B of][]{2002A&A...383..423E} where the synchrotron kernel is
  replaced by a delta distribution, $\delta(\nu-\nus)$ with $\nus = 3
  e B \sin\theta\, \gamma^2 / (2\pi m_\rmn{e} c)$, where $\theta$ is
  the CRes' pitch angle and which gives the exact synchrotron formula
  for a power-law electron population with spectral index $\alpha=3$
  and attains only order unity corrections ($<20$ per cent) for small
  spectral changes $\Delta \alpha<0.5$.} is $\sim 3 \gamma^{2}
\nu_\rmn{c}$, where $\nu_\rmn{c}$ is the non-relativistic cyclotron
frequency; this implies that for a given observation frequency $\nus$,
the greatest contribution comes from electrons with
\begin{equation}
  p_{\rm emit} \approx \gamma_{\rm emit} \approx 5 \times 10^{3} 
  \left( \frac{\nus}{1 \, {\rm GHz}} \right)^{1/2} 
  \left( \frac{B}{5 \, \mu{\rm G}} \right)^{-1/2} (1+z)^{1/2}, 
  \label{eqn:pemit} 
\end{equation}
where for instance $B \approx 5 \, \mu$G has been inferred from high
resolution measurements of spectral aging in the sausage relic
\citep{2010Sci...330..347V}. Figure~\ref{fig:p_contrib_reaccel}
illustrates the contribution per logarithmic interval to the integral
in equation~\eqref{eqn:reaccel}, for $p_{\rm emit}=10^{4}$. For a
strong shock, we see that the integrand has a broad peak ranging from
$1 \lsim p^{\prime} \lsim 100$, while for a weak shock it increases
monotonically with $p^{\prime}$, receiving its dominant contribution
from $10^{2} \lsim p^{\prime} \lsim 10^{4}$.

We can understand this as follows. The integrand peaks when $\alpha -
\beta(p^{\prime}) = 0$, where $\beta(p^{\prime}) = -d\log \, f_{0}/
d\log \, p^{\prime}$, i.e. when the power-law slope of reaccelerated
particles and the fossil distribution function have the same slope. It
picks up most of its contributions from a neighborhood of this
region. From Fig.~\ref{fig:e_spec_represent}, we see that
$\beta(p^{\prime})$ ranges from zero (due to NR Coulomb cooling) to
extremely steep positive slopes $\beta(p^{\prime}) > 3$ (due to
inverse Compton and synchrotron cooling), this tangent is guaranteed
to exist. In practice, the relevant regime where $2 <
\beta(p^{\prime}) < 3$ lies in $1 \lsim p^{\prime} \lsim 10^{3}$, the
'adiabatic' regime which only only mildly affected by relativistic
Coulomb cooling or inverse Compton cooling. Regions which have much
shorter cooling times and consequently had their injected slopes
strongly modified do not contribute significantly. This suggests the
fortunate conclusion that we do not need to accurately resolve regions
strongly affected by cooling to accurately predict the reaccelerated
spectrum. An important point to note is that in this crucial adiabatic
momentum regime $1 \lsim p \lsim 100$, the spectral index of the
injected spectrum $\alpha \approx 2.5$ (e.g., see bottom panel of
Fig.~\ref{fig:e_spec_represent}), indicating that weak shocks ($\mach
\sim 3$) dominate the assembly of the fossil CRe spectrum here. We
will return to this point in \S\ref{sect:shock_model_vary}.

\subsection{Do Our Simulations Have Sufficient Time Resolution?} 
\label{sect:time_resolution}
\begin{figure*}
\begin{minipage}{2.0\columnwidth}
  \includegraphics[width=0.5\columnwidth]{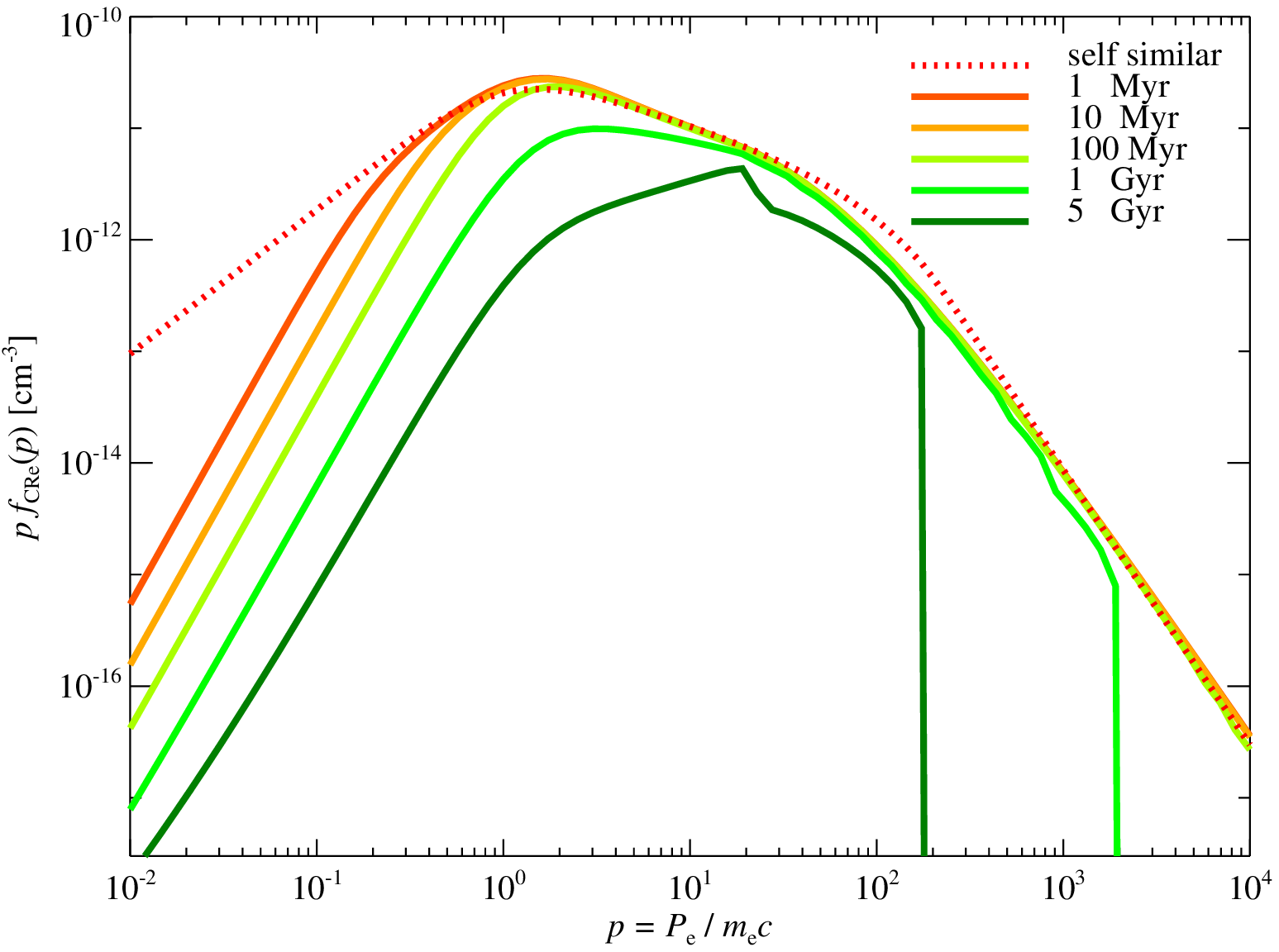}
  \includegraphics[width=0.5\columnwidth]{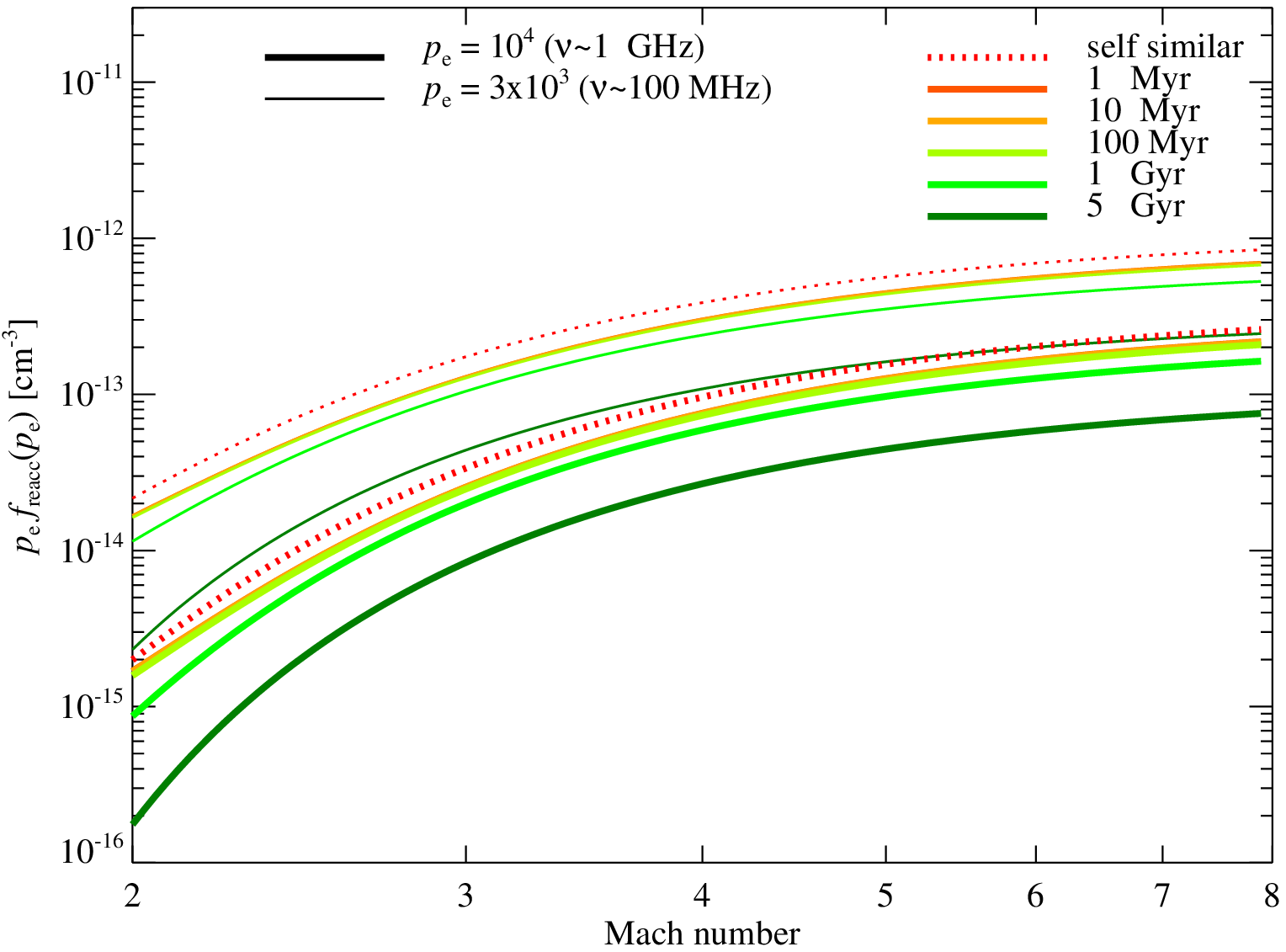}
  \caption{Continuous versus discrete injection. We show the cooled
    fossil CR electron distribution function as a function of
    normalized momentum. The self similar distribution (red dotted)
    assumes constant continuous injection and constant density and is
    compared to discrete injection scenarios of CRes with different
    duty cycles: 1~Myr (dark orange), 10~Myr (light orange), 100~Myr
    (light green), 1~Gyr (green), 5~Gyr (dark green). We assume an
    injected CR electron population given by
    equation~\eqref{eq:Qinj_spectra} and an electron number density,
    $n_\e=3\times10^{-5}\,\cm^{-3}$. The distribution function from
    discrete injection approaches the self similar solution when the
    injection timescale becomes smaller than the cooling time. Note
    that for this figure we adopt a characteristic assembly time of
    the CRes of 10~Gyrs, hence the 5~Gyrs curve (at $p=10$) and the
    1~Gyr curve (at $p=10^3$) are superpositions of two and ten
    injected spectra, respectively, which explains the visible
    shoulders for these curves.}
\label{fig:time_resolve_analytic} 
\end{minipage}
\end{figure*}

We now turn directly to the question of required time resolution. Our
finding in \S\ref{sect:analytic} that injection and cooling is roughly
time steady appears to contradict the approach we have taken in our
simulations, where we inject CRes in discrete bursts (due to finite
time resolution). In Fig.~\ref{fig:time_resolve_analytic}, we show the
distribution function obtained by mimicking the same procedure we
followed in our simulations: injecting the electrons in discrete
bursts, and then passively cooling the injected population from each
time step, via equations~\eqref{eq:f_evolv}-\eqref{eq:f_evolv_diff},
for different values of $\Delta t$ (in our simulations, $\Delta t =
(100\,\rmn{Myrs} - 1.5\, \rmn{Gyrs})$, see
Table~\ref{tab:cluster_sample} for details). For a fair comparison, we
have assumed that cooling and injection are exactly constant. By
comparison with the left panel in
Fig.~\ref{fig:time_resolve_analytic}, we see that in regions where
$\Delta t < t_{\rm cool}$, our computational procedure correctly
approaches the analytic solution. However, when $\Delta t > t_{\rm
  cool}$, the discrete grid overestimates the importance of cooling,
causing the numerical solution to fall below the analytic
one.

However, our results from the previous section suggest that the
regions where cooling is important do not contribute significantly to
$f(p_{\rm emit})$. In particular, the main important mechanism is
non-relativistic Coulomb cooling at $p \lsim 1$. This momentum regime
of the fossil spectrum does not contribute significantly to $f(p_{\rm
  emit})$ (as in Fig. \ref{fig:p_contrib_reaccel}) because it is 'too
far from the action' at $p_{\rm emit}$, and also has been
significantly depopulated. Thus, our finite time resolution does not
significantly affect results. We show this explicitly in the right
panel in Fig.~\ref{fig:time_resolve_analytic}, which shows $p_{\rm
  emit}\,f_{\rm reacc}(p_{\rm emit})$ from the reaccelerated
population for the various curves in the left panel of
Fig.~\ref{fig:time_resolve_analytic}, as a function of shock Mach
number. The differences are small, given other uncertainties in the
problem. This shows that the contribution from fossil reacceleration
can indeed be estimated accurately analytically. Simulations are
nonetheless invaluable for their ability to shed light on spatial (see
the insets of Fig.~\ref{fig:density_evol}) and temporal variations
(see \S\ref{sect:cluster_vary}) in injection and cooling rates, all of
which can lead to significant scatter in the CRe fossil population.

\subsection{Principal Result: Reacceleration Dominates over Direct Injection}

\begin{figure}
  \includegraphics[width=1.0\columnwidth]{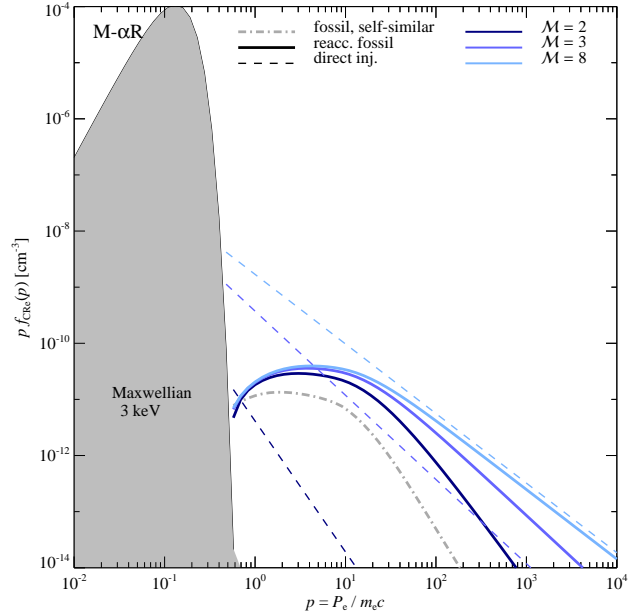}
  \caption{Median $p f_{\rm CRe}$ for direct injection (dashed line)
    and reacceleration (solid line), for different Mach numbers (color
    coded); also shown is the fossil distribution function
    (dot-dashed). $f_{\rm CRe}$ for direct injection is a power-law
    attached to the Maxwellian, while the reaccelerated $f_\rmn{CRe}$
    is a boosted version of the fossil distribution function, with an
    asymptotic power-law tail. The distribution function falls much
    more drastically at low Mach number for direct injection (dashed)
    than for reacceleration (solid).}
  \label{fig:e_spec_all_reacc}
\end{figure}
    
Thus far, we have focused on accurate calculations of the fossil
electron distribution function. We now explicitly illustrate, using
equation~\eqref{eqn:reaccel}, how this population of electrons is
transformed and boosted by a shock. Figure~\ref{fig:e_spec_all_reacc}
shows the initial fossil electron distribution function and the
transformed distribution function after Mach number ${\mathcal
  M}=2,3,8$ shocks; the population due to direct injection is also
shown. The distribution function for direct injection is a power-law
attached to the Maxwellian, while the reaccelerated distribution
function is a boosted version of the fossil distribution function,
with an asymptotic power-law tail. This tail has the same power-law
slope $\alpha$ as for direct injection. Note that while the total
number of fossil electrons is conserved, the number {\it density} of
reaccelerated electrons increases over that of the fossil population
by the shock compression factor $r$, which explains the normalization
boost upon reacceleration. It is clear that while the electrons from
direct injection have a strong Mach number dependence, this trend is
much weaker for the reaccelerated population. This suggests that
reacceleration could dominate at low Mach numbers.

\begin{figure}
  \centering
  \includegraphics[width=1.0\columnwidth]{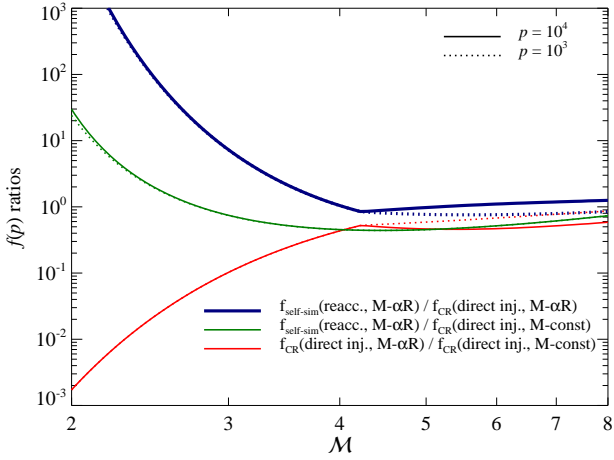}
  \caption{The ratio of the distribution function evaluated at
    $p\equiv P_\e/\mec=10^3$ (dotted) and $10^4$ (solid), for
    different acceleration models. The fiducial dark blue curves show
    $f_{\rm CR}({\rm reacc.})/f_{\rm CR}({\rm inj})$. Reacceleration
    is roughly comparable to direct injection at high Mach numbers and
    vastly exceeds direct injection at low Mach numbers. The green
    (red) curves compare reacceleration (direct injection) with a
    strawman model--commonly used in the literature--where a constant
    fraction $\zeta_{\rm inj,e}=0.01$ of the shock energy goes into
    electrons, independent of Mach number. The straw man model has
    roughly similar behavior to the other models at high Mach number,
    but vastly divergent behavior at low Mach number. Note that the
    reaccelerated CRe fossil population, $f_{\rm CR}({\rm reacc.})$,
    is derived from the self-similar model in
    equation~\eqref{eq:f_self_similar} using a characteristic cooling
    time $\Delta t\sim9$~Gyrs.}
\noindent \hrulefill
 \label{fig:f_ratio}
\end{figure}

Figure~\ref{fig:f_ratio} shows the main result of this paper, the
ratio $f_{\rm reacc}(p_{\rm emit})/f_{\rm inj}(p_{\rm emit})$, or the
ratio of the distribution function at the primary emitting frequency
for the reaccelerated population to the same for the freshly injected
population (dark blue curve). We see that at high Mach numbers
(${\mathcal M} > 4$), relic reacceleration and fresh injection are
comparable, but for low Mach numbers (${\mathcal M} < 4$), fossil
reacceleration vastly dominates over fresh injection. We also compare
the reacceleration and direct injection models to a straw man model
(`M-const') where a constant fraction of the shock energy $\zeta_{\rm
  inj,e}=0.01$ is converted to relativistic electrons, independent of
Mach number. As previously mentioned, while we do not believe this
assumption to be physically realistic, it has been widely used in the
literature, and gives us a baseline for comparison. While the
`M-const' model gives comparable results to the other two models for
${\mathcal M} > 4$, it differs sharply from the direct injection model
for ${\mathcal M} < 4$, {\it over-predicting} the CRe population. This
makes sense, since physically we expect the injection efficiency to
plummet at low Mach numbers. By contrast, the `M-const' model is
similar to the reacceleration model down to ${\mathcal M} \sim 3$, but
for ${\mathcal M} < 3$ it {\it under-predicts} the CRe
population. This has two important consequences. Firstly, it means
that `M-const' models have been getting the right answer for the wrong
reasons: while they can match observations of ${\mathcal M} \gsim 3$
relics, this is not necessarily because acceleration efficiency is
independent of Mach number. {\it Instead, our model provides a
  physical basis for the observed brightness of low Mach number
  relics, through the existence of fossil CRe} (see also
Fig.~\ref{fig:Fluc_Mach}). Secondly, at extremely low Mach numbers
${\mathcal M} < 3$, reacceleration predicts many more CRe than even
this model. Thus, {\it we predict many more relics with steep spectra
  which are potentially observable with LOFAR} (see
\S~\ref{sec:flux}).

How can we understand the Mach number dependence of $f_{\rm
  reacc}(p_{\rm emit})/f_{\rm inj}(p_{\rm emit})$? It is easy to
understand why it rises steeply at low Mach numbers. The Mach number
dependence of $f_{\rm reacc}(p_{\rm emit})$ is solely due to the
dependence of the power-law slope $\alpha$ on Mach number; the number
of particles available for acceleration is fixed. As $\alpha$ steepens
at low Mach number, $f_{\rm reacc}(p_{\rm emit})$ falls in a power-law
fashion. This mostly happens for ${\mathcal M} \lsim 4$, when $\alpha$
begins to evolve significantly. On the other hand, fresh injection is
affected both by the change in $\alpha$ and more importantly, the
change in the number of particles available for acceleration. As can
be easily derived from the jump conditions, low Mach number shocks
have lower thermalization efficiencies: the post-shock gas has a lower
temperature and higher bulk velocity. This means that $x_{\rm inj} =
p_{\rm inj}/p_{\rm therm}$ has to be higher for a particle to be able
to overcome the bulk fluid motion to cross the shock. This increase in
$x_{\rm inj}$ on the Maxwellian tail exponentially reduces the number
of particles which are accelerated at low Mach number shocks, which is
why $f_{\rm inj}(p_{\rm emit})$ is exponentially suppressed.
  
On the other hand, the fact that $f_{\rm reacc}(p_{\rm emit})$ and
$f_{\rm inj}(p_{\rm emit})$ are roughly comparable at high Mach
numbers may appear somewhat surprising. Since the build-up of the CRe
fossil population is due to the interaction of continuous injection
via multiple shocks with cooling processes, a priori we might expect
that it could potentially be orders of magnitude smaller or larger
than that due to direct injection at a single shock. In particular, as
we have seen in Fig.~\ref{fig:p_contrib_reaccel}, since much of the
contribution to $f_{\rm reacc}(p_{\rm emit})$ comes from an adiabatic
regime which grows monotonically on cosmological timescales, we might
expect the reservoir of fossil CRe built up to outweigh the
contribution from a single shock. In fact, as seen in
Fig.~\ref{fig:e_specZ} and Fig.~\ref{fig:density_evol}, the injection
rate is fairly time-steady, with late times having a somewhat larger
contribution since the gas is getting hotter, increasing $p_{\rm
  therm}$. The characteristic injection timescale is $t_{\rm dyn} \sim
t_{\rm H}/\sqrt{\Delta} \sim 1$ Gyr, implying that the fossil CR
electron population should be $t_{\rm H}/t_{\rm dyn} \sim 10$ times
larger. The fact that cooling operates, and the reduced contribution
from early times makes the value of $f_{\rm reacc}(p_{\rm
  emit})/f_{\rm inj}(p_{\rm emit}) \sim 1$ shown in
Fig.~\ref{fig:f_ratio} reasonable. We also note that the fossil CRe
population can show significant scatter from cluster to cluster, up to
an order of magnitude (which we shall discuss in
\S\ref{sect:cluster_vary}), so one should not over-interpret the exact
value of $f_{\rm reacc}(p_{\rm emit})/f_{\rm inj}(p_{\rm emit})$.

\section{Variations in the Fossil Electron Spectrum} 
\label{sect:variations}

\subsection{Dependence on Shock Acceleration Model} 
\label{sect:shock_model_vary} 

In \S\ref{sec:models}, we discussed 3 different models for particle
acceleration, but thus far largely used one model (`M-$\alpha$R') to
compute the fossil electron spectrum. Fig.~\ref{fig:f_ratio} has
already highlighted differences between the `M-$\alpha$R' model and
the `M-const' model for a single shock. We now propagate these
differences between the three models when the entire assembly history
of the fossil population is taken into account.

\begin{figure}
  \includegraphics[width=1.0\columnwidth]{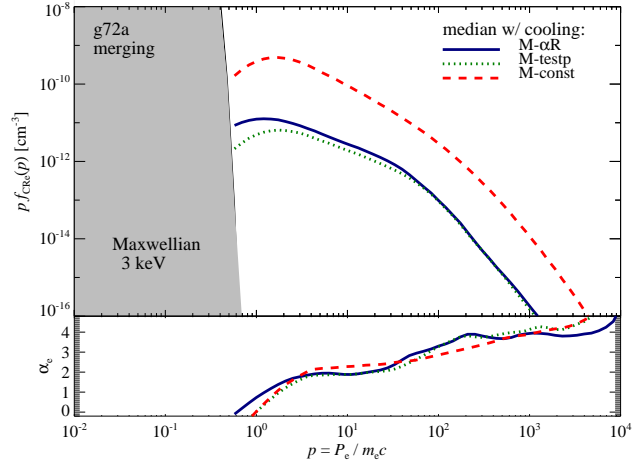}
  \caption{Comparison of median CRe spectra in the cluster outskirts
    of g72a at $z=0$ in different acceleration models. Shown are the
    test particle model (`M-testp'), the CRe spectral index
    renormalization model (`M-$\alpha$R'), and the model with constant
    acceleration efficiency $\zeta_{\rm inj,e}=0.01$ (`M-const').}
  \label{fig:e_spec_CRmodel}
\end{figure}

The resulting fossil electron population for the test particle model
(`M-testp'), the CRe spectral index renormalization model
(`M-$\alpha$R'), and the model with constant acceleration efficiency
$\zeta_{\rm inj,e}=1$ per cent (`M-const') is shown in
Fig.~\ref{fig:e_spec_CRmodel}. We see that while the `M-$\alpha$R' and
'M-testp' models give very similar results, the `M-const' model yields
a fossil distribution function with a significantly higher
normalization.

These results are easy to understand. From the bottom panel of
Fig.~\ref{fig:e_spec_represent} (and discussed in
\S\ref{sec:momentum_regime}), we see that the spectral index of the
median injected population before cooling is $\alpha_{e} \sim 2.5$,
i.e., weak shocks with ${\mathcal M} \sim 3$ dominate the assembly of
the CRe fossil population. The `M-$\alpha$R and `M-testp' models are
identical at low Mach numbers; they only differ in how energy
conservation is enforced at high Mach numbers (even so, differences
are slight). By contrast, the `M-const' model is broadly similar to
the other models at high Mach numbers, but diverges sharply at low
Mach numbers, with a significantly higher normalization (c.f. the red
curve in Fig.~\ref{fig:f_ratio}), since by construction $\zeta_{\rm e,
  inj}$ is constant, whereas it plummets drastically at low ${\mathcal
  M}$ for the other models. Since fossil CRes are primarily put in
place by weak shocks, it is not surprising that they are much more
abundant in this model. This amounts then to an increase in
normalization of the `M-const' model by roughly an order of magnitude
in comparison to the other models (as can be inferred for the
acceleration efficiency at $\mach=3$ shocks from Figs.~\ref{fig:f_ratio}
and \ref{fig:etamax}, considering that the relevant shocks are
preferentially those at late time in a hot medium with particle
energies of a few keV).

These results indicate our main conclusion that fossil reacceleration
must dominate at low Mach number is {\it independent} of the shock
acceleration model. If one attempts to explain the observed brightness
of weak shocks by modifying shock physics and increasing the
acceleration efficiencies at low ${\mathcal M}$, the same would hold
for previous weak shocks and the normalization of the fossil CRe
population would be correspondingly larger. Since the shock
acceleration efficiency largely cancels out, the relative importance
of reacceleration vs. direct injection depends on accretion/merger
history and cooling, for which (unlike shock acceleration) there is
relatively little uncertainty in the physics.

\subsection{Cluster-to-Cluster variations} 
\label{sect:cluster_vary} 
\begin{figure*}
\begin{minipage}{2.0\columnwidth}
  \includegraphics[width=\columnwidth]{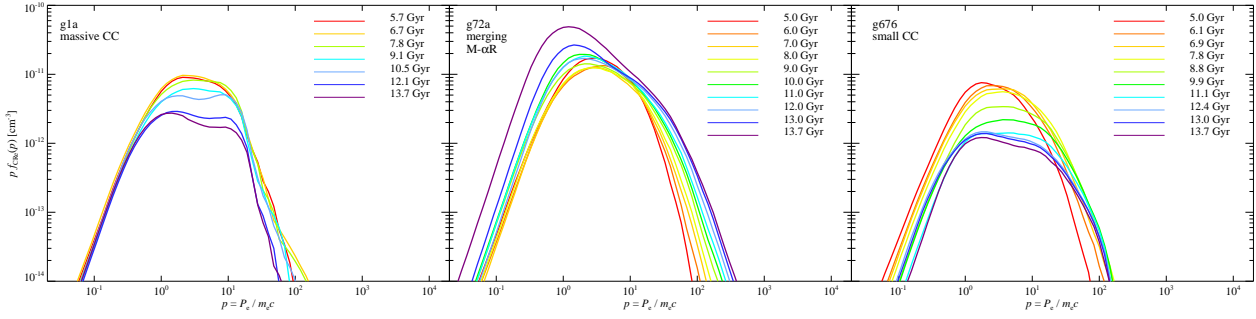}
  \caption{Lagrangian evolution of CRe spectra of three different
    clusters.  We show the median CRe distribution function of a
    representative sample of SPH particles that all end up at $z=0$ in
    the virial regions within $(0.8-1.0)\,\rvir$ (different time
    intervals since Big Bang are color coded). Shown are a large cool
    core cluster (g1a, left), a cluster with a recent merger (g72a,
    middle), and a small cool core cluster (g676, right). There is
    strong evolution of $f_\rmn{CRe}$ in the post-merger cluster,
    which is noticeably reduced in the cool core clusters. While the
    normalization is increasing in our post-merger cluster that
    experienced a more violent recent history, the opposite trend is
    visible in the cool core clusters where CRe cooling processes
    dominate. However, the overall shape of the distribution function
    is very similar between those clusters that vary widely in
    dynamical stage and mass.}
 \label{fig:e_spec_z}
\end{minipage}
\end{figure*}

{\it Is the build-up of a CRe spectrum in cluster outskirts a
  universal process?} To answer this question, we compare the CRe
distribution for three typical clusters at different times in
Fig.~\ref{fig:e_spec_z}: a small cool core cluster, a large cool core
cluster, and a large post-merger cluster. It is remarkable how similar
the spectral shape is between the different clusters and at different
times. This generic shape can be understood from the analytic model
presented in \S\ref{sect:analytic}. The peak is at $p \sim 1$ and the
normalization for each cluster changes by a factor less than 10. The
recent merger cluster g72a shows the highest CRe number density, while
the CRes in the two cool core clusters have a lower abundance.

\begin{figure}
  \includegraphics[width=1.0\columnwidth]{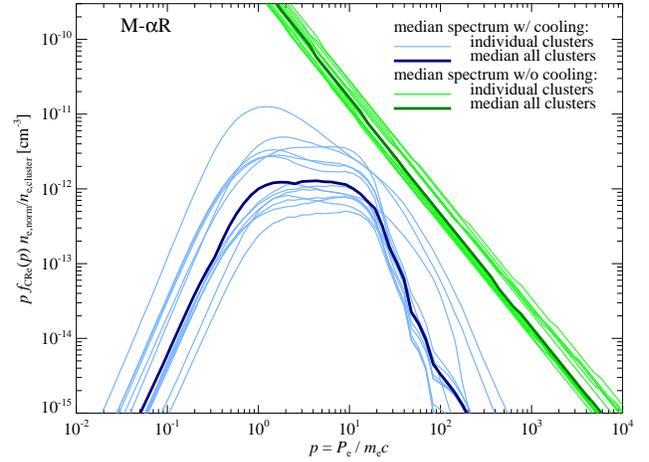}
  \caption{CRe spectra in the virial regions of clusters at $z=0$.
    Shown are the median CRe spectra of all simulated clusters
    normalized to the same thermal electron number density
    $n_\rmn{e,norm}=3\times10^{-5}\, \cm^{-3}$; with cooling (light blue) and
    the median of those (dark blue), injected spectra (light green)
    and the median of those (dark green). The injected spectrum
    appears quite universal with little scatter. By contrast, although
    the cooled spectra all have similar shapes, their normalizations
    can differ substantially, by up to a factor of $\sim 50$.}
\label{fig:e_spec_all}
\end{figure}

What is the origin of this dispersion? Does it originate from
differences in CRe injection or cooling, between clusters? In
Fig.~\ref{fig:e_spec_all}, we show the CRe spectra for all 14
clusters, both with (blue) and without (green) cooling. The adiabatic
spectra (which were only transported adiabatically without accounting
for cooling) exhibit remarkably little scatter, while the cooled
spectra show considerably more scatter. This suggests that the
variance between clusters arises from the cooling rather than the
injection process (for more on scatter in the adiabatic spectrum, see
Appendix~\ref{sect:CRinjection} and Fig.~\ref{fig:Qinj_fit}).

Two other features in Fig.~\ref{fig:e_spec_all} are worth
noting. Firstly, cooling introduces increased scatter even in the
ostensibly quasi-adiabatic regime $10 \lsim p \lsim 100$. Note that
this regime has $t_{\rm cool} > t_{\rm H}$ only for $n_{e} \lsim
10^{-5} \, {\rm cm^{-3}}$ and at $z=0$. For higher gas densities and
redshifts, some cooling is possible; thus, variations in gas clumping
(which affects the amount of Coulomb cooling) and shock history (which
affects the total amount of cooling; cf. the difference between
``impulsive'' and ``continuous injection'' solutions). The fact that
there is substantial scatter at $p \sim 100$ (where only inverse
Compton cooling is important) points toward the latter. Namely,
clusters which receive their dose of CRes earlier than others undergo
more inverse Compton cooling. The different build-up and normalization
of the CRe population in the cool core and recent merger clusters in
Fig.~\ref{fig:e_spec_z} also supports this. Secondly, the scatter at
low $p$ is considerably higher than at high $p$. Part of this is
because the cooling time is shorter at low $p$, and hence differences
in shock history--which are averaged over shorter timescales--are
amplified. However, part of the difference is artificial, and due to
our finite time resolution, which overestimates the effects of cooling
(\S\ref{sect:time_resolution} and
Fig.~\ref{fig:time_resolve_analytic}). For instance, the two clusters
with the higher normalizations in the Coulomb cooling regime are g72a
and g72b, which have a time resolution ten times better than the other
clusters. Fortunately, as shown in \S\ref{sec:momentum_regime}, this
momentum regime is unimportant in its contribution to the
reaccelerated spectrum at $p_{\rm emit}$. By contrast, the cooling
time is generally longer than our time resolution in the $p \gsim 10$
regime.

\begin{figure}
  \includegraphics[width=1.0\columnwidth]{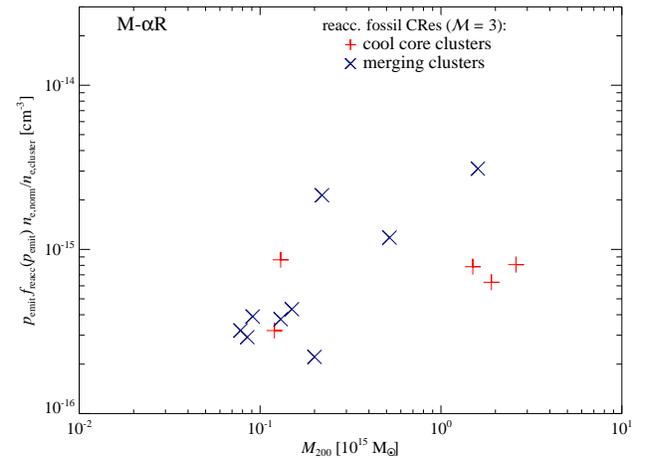}
  \caption{The momentum integrated quantity $p_{\rm emit} f(p_{\rm
      emit})$ (where $p_{\rm emit} \sim 10^{4}$) plotted against
    cluster mass. For each cluster the CRes are normalized to the same
    thermal electron number density $n_\rmn{e,norm}=3\times10^{-5}\,
    \cm^{-3}$. The red crosses and blue $X$ show the simulated cool
    core (CC) and merging clusters,
    respectively. \label{fig:mass-scaling}}
\end{figure}

Finally, by integrating over momentum (equation~\ref{eqn:reaccel}), we
show the scatter in the observationally more relevant quantity $p_{\rm
  emit} f(p_{\rm emit})$, where $p_{\rm emit} \sim 10^{4}$, in
Fig.~\ref{fig:mass-scaling}, plotted against cluster mass $M_{200}$.
The scatter is about an order of magnitude, and somewhat below the
apparent scatter in Fig.~\ref{fig:e_spec_all}, since it is weighted
towards the lower scatter, high $p$ regime. There may be a trend for
the CRe distribution function to increase with $M_{200}$ for merging
clusters, though our sample size is too small to make definitive
statements.

\subsection{Radial variations of the fossil electron spectrum}
\label{app:radial}

\begin{figure*}
  \begin{minipage}{2.0\columnwidth}
    \includegraphics[width=0.5\columnwidth]{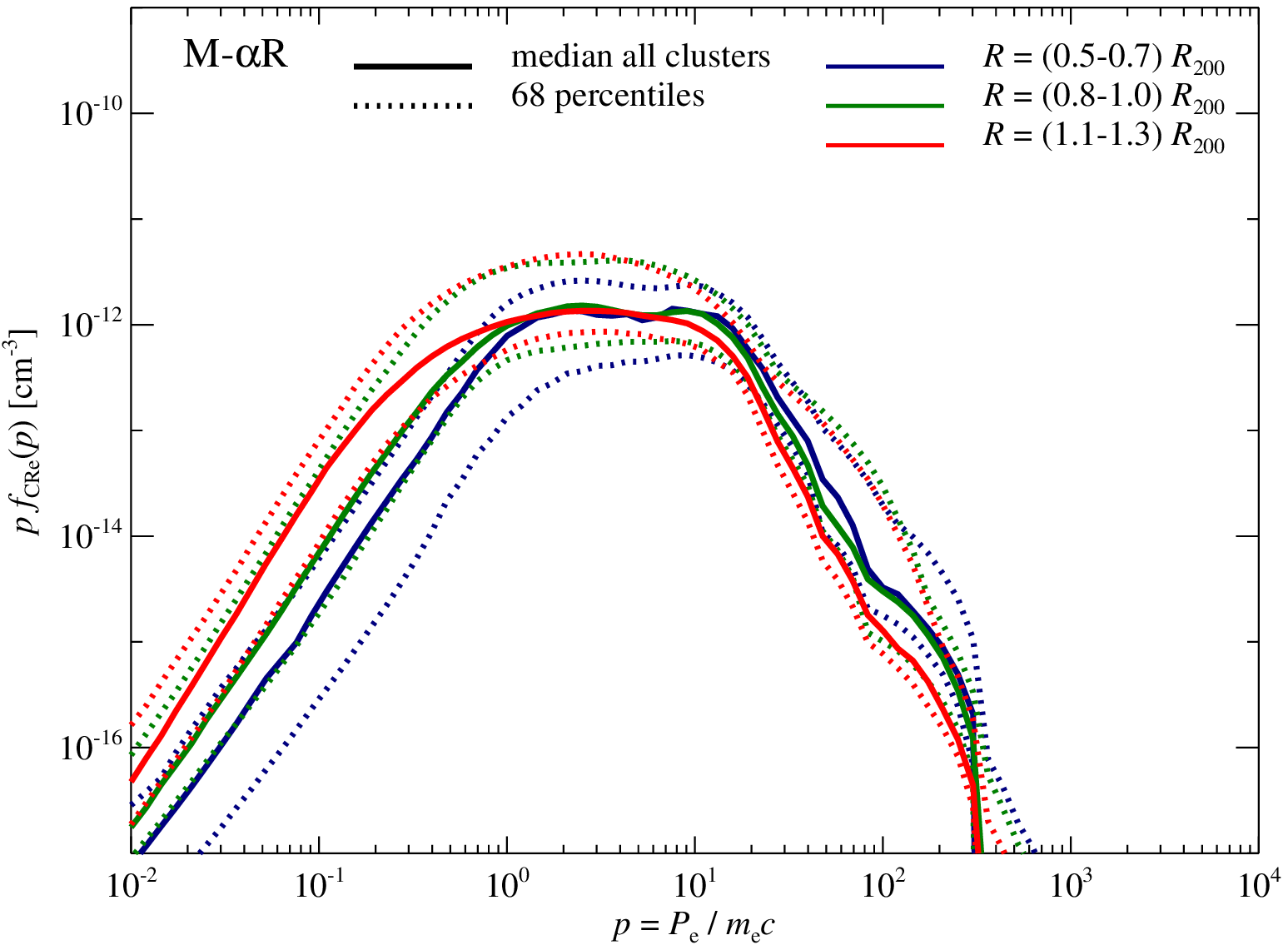}
    \includegraphics[width=0.5\columnwidth]{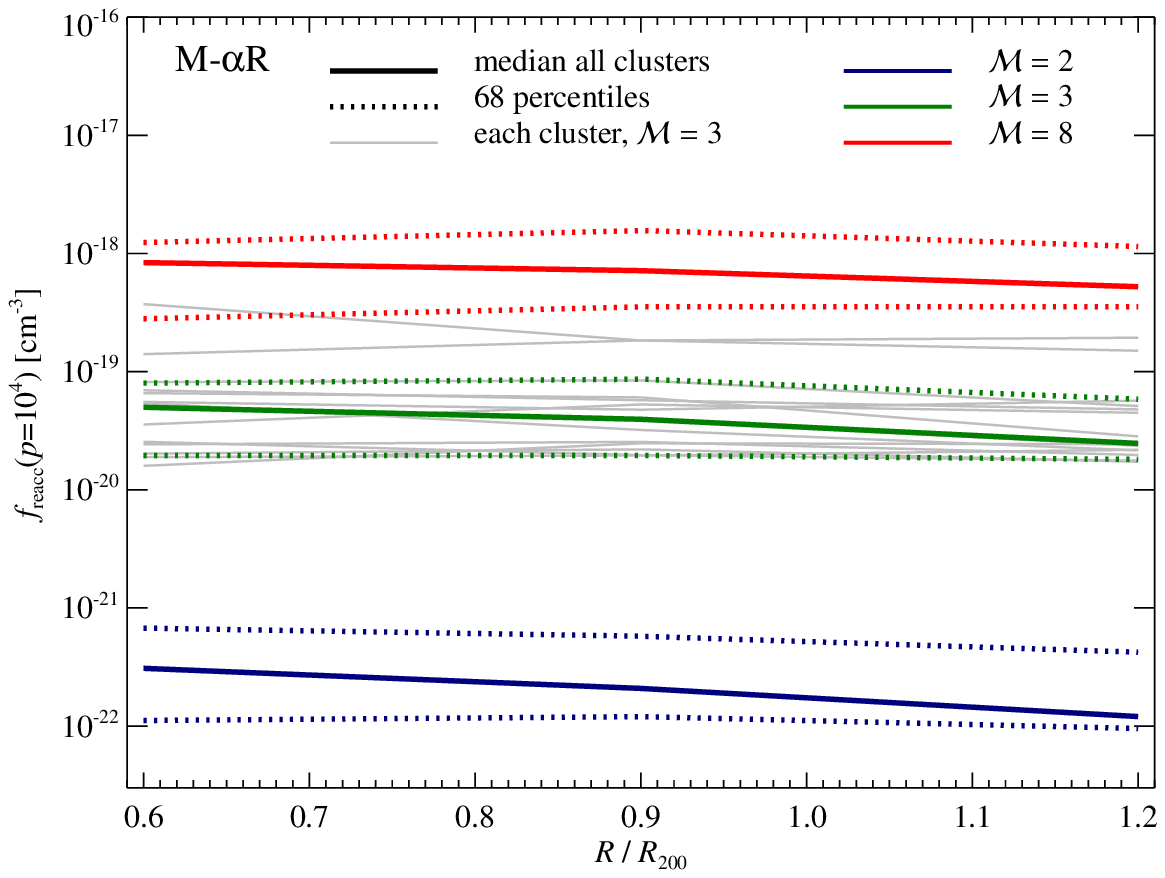}
    \caption{Radial variations of the fossil CRe spectrum in different
      clusters. We show the CRe distribution function, where the solid
      lines represent the median of all clusters, the dotted lines the
      68 percentiles. In the left panel we show the {\em fossil}
      distribution function as a function of momentum where the colors
      represent different radial bins: blue line $(0.5-0.7)\,\rvir$,
      green line $(0.8-1.0)\,\rvir$, and red line $(1.1-1.3)\,\rvir$.
      In the right panel we show the {\em reaccelerated} distribution
      function at momentum $p=10^4$ as a function of radius where the
      colors represent different Mach numbers of the reaccelerating
      shock: $\mach=2$ (blue), $\mach=3$ (green), and $\mach=8$
      (red). The thin grey lines show the reaccelerated CRe spectra
      for a $\mach=3$ shock in each individual cluster. The
      reaccelerated spectra at $p=10^4$ show only a very weak radial
      trend.}
    \label{fig:f_radial}
  \end{minipage} 
\end{figure*}

To quantify the potential bias from the uncertainty in the distance of
relics to the cluster center, we explore in Fig.~\ref{fig:f_radial}
the radial variations of both the fossil CRe spectrum and CRes that
are reaccelerated in typical merger shocks. To this end, we probe
three radial bins; $(0.5-0.7)\,\rvir \sim R_{500}$, $(0.8-1.0)\,\rvir$
(the main focus of this paper), and $(1.1-1.3)\,\rvir \sim
R_\rmn{vir}$. We find that the distribution of fossil CRes in the
non-relativistic Coulomb cooling regime and high energy IC regime is
sensitive to the radius. The variation at low momenta (due to the
difference in Coulomb cooling; the density increases inward by a
factor of 2-3 in each radial bin) is larger; variations at high
momenta (due to differences in IC cooling, which is density
independent and only varies with the relative injection history) are
smaller. In contrast, at intermediate momenta, $p\sim 1-10$, there are
almost no differences between the median distribution functions of our
cluster sample. We have already seen such flat radial variations in
CRp profiles \citep{pinzke10} (note that CRp's are effectively
adiabatic due to inefficient cooling). Indeed, in the hadronic model,
a flat inferred CRp profile is required to explain the surface
brightness profile in giant radio halos such as Coma
\citep{2012arXiv1207.6410Z}. The fact that the CR number density is
flat with radius implies that the relative number density of CRs $\eta
= n_{\rm CRe}(> p^{\prime})/n_{\rm th}$ {\it increases} with
radius. This may seems surprising, given that the cluster outskirts
are generally assembled via weaker shocks, due to the slowdown of
structure formation in a $\Lambda$CDM universe. As discussed in
\S\ref{sect:sim_spectrum}, this effect arises because the distribution
function from later shocks have higher normalization: they take place
in a hotter medium, so that $p_{\rm inj} \propto T^{1/2}$ attaches to
the thermal distribution at higher momenta. This effect is apparently
sufficiently to render $n_{\rm CRe}(> p^{\prime}) \sim$ const.

The spectral shape of the fossil CRe distribution functions is similar
between clusters, however, the normalizations can differ up to a
factor 100. The spectral shape is also similar between different
radial bins, indicating similar injection and cooling histories. In
the right panel of Fig.~\ref{fig:f_radial} we explore the
reaccelerated fossil population at $p=10^4$ for different Mach numbers
of merger shocks. We find an increasing abundance of reaccelerated
CRes for smaller radii. The reason is that the reaccelerated CRe
populations for $\mach>2$ shocks, are mainly build up from the CRes
with $p>10$, hence we expect a factor $1-3$ difference between the
different radii.

\section{Comparison with Radio Relic Observations}
\label{sec:flux}

\begin{table*}
\begin{minipage}{2.0\columnwidth}
\caption{Relic sample. For simplicity we assume a downstream magnetic
  field of 5~$\mug$ and a downstream temperature of $kT_2 = 5$~keV.}
\begin{tabular}{l c c c c c c l}
\hline
\hline
relic name & redshift & $\mach^{(1)}$ & $\Psi_\rmn{S}^{(2)}$ & thickness$^{(3)}$ & $n_\e^{(4)}$ & $P$(1.4~GHz) & reference\\
& & & [Mpc$^2$] & [kpc] &[el cm$^{-3}$]& [$10^{32}$ erg s$^{-1}$ Hz$^{-1}$] & \\
\hline
A 2256     & 0.0594 & 2.6 & 0.6 & 70 & $\,\,10^{-3}$     & 0.4 & \citet{2011MmSAI..82..547C} \\
A 3667     & 0.055  & 4.7 & 2.0 & 62 & $8\times 10^{-5}$ & 4.1 & \citet{1997MNRAS.290..577R} \\
Sausage    & 0.1921 & 4.5 & 1.5 & 48 & $3\times 10^{-5}$ & 1.4 & \citet{2010Sci...330..347V} \\
Toothbrush & 0.225  & 4.6 & 3.5 & 45 & $\,\,10^{-4}$     & 6.0 & \citet{2012AA...546A.124V} \\
A 2744     & 0.3080 & 2.4 & 2.6 & 44 & $7\times 10^{-4}$ & 0.5 & \citet{2007AA...467..943O} \\
\hline 

\end{tabular}  \begin{quote}
 Notes:\\ (1) The Mach number $\mach$ is derived using the observed
 spectral index of the radio emission closest to the shock front. (2)
 Shock area, estimated from the largest linear size of relic squared.
 Note that there is a large uncertainty of the relic in the direction
 of the line of sight. (3) Relic thickness as estimated from the
 cooling length, equation~\eqref{eq:l_cool}. (4) Assumed electron
 density adjusted to match observed radio luminosities.
\end{quote}
\label{tab:relic_sample}
\end{minipage}
\end{table*} 

In this section we derive the radio synchrotron emission from our
fiducial 'M-$\alpha$R' CRe model and compare it to observations of
radio relics. These comparisons assume the relics lie between
$(0.8-1.0)\,\rvir$ and explore a limited range of parameter space; they
are only meant to be illustrative.  Our main task is to explore
distinctive observational signatures of fossil electrons. As expected,
low Mach number shocks are much brighter if fossil electrons abound;
we therefore predict many more steep spectrum sources to be detectable
at low flux limits. On the other hand, the relic luminosity function
is {\it not} a robust discriminant of models with and without fossil
electrons. Thus, spectral information is needed to test our model.

Given the many uncertainties, we adopt a simple model to estimate
radio luminosities, which still takes the effects of cooling into
account. The radio synchrotron emissivity of a power-law distribution
of CRes is \citep{1979rpa..book.....R}:
\begin{eqnarray}
\label{eq:synch_emis}
J(\nu) &\approx& J_0\,C_\rmn{reacc}
\Gamma\left(\frac{3\,\alpha-1}{12}\right)\,
\Gamma\left(\frac{3\,\alpha+19}{12}\right)\,
\left(\frac{\nu}{\nu_c}\right)^{-\alpha_\nu}\,,\nonumber \\
&& \rmn{where}\quad J_0 \equiv \frac{3^\frac{\alpha}{2}\,e^2\,\nu_c}{c\,\left(\alpha+1\right)}\,.
\end{eqnarray}
Here the radio spectral index $\alpha_\nu=(\alpha-1)/2$, $\alpha$ is
the spectral index of the CRe population, $\Gamma$ is the gamma
function, and $\nu_{c}=eB/(2 \pi m_{\rm e} c)$ is the cyclotron
frequency. The $\sim 10$ GeV electrons which emit at $\sim$GHz
frequencies (equation~\ref{eqn:pemit}) cool via IC and synchrotron
emission over a post-shock distance:
\begin{eqnarray}
  \label{eq:l_cool}
  l_{\rm cool} \approx u_{2} t_{\rm cool} &\approx& 200 \, {\rm kpc} 
\left(\frac{u_{2}}{10^3\,\rmn{km}\,\rmn{s}^{-1}}\right)
\,\left(\frac{\nus}{1 \, {\rm GHz}} \right)^{-1/2}\nonumber \\
  &\times&\left(\frac{B_{2}}{5\,\mug}\right)^{1/2}
\left(\frac{B^2_{\rm eff, 2}}{\left(5\,\mug\right)^2}\right)^{-1}(1+z)^{-1/2}\,,\nonumber \\
  &&
\end{eqnarray}
where $B_{\rm eff,2}^{2}= B_{2}^{2} + B_{\rm CMB}^{2}$, $B_{\rm
  CMB}=3.24 (1+z)^{2} \, \mu$G, and typical downstream velocities,
$u_2$, in a relic ~$10^2-10^3\,\rmn{km}\,\rmn{s}^{-1}$. The specific
luminosity is given by:
\begin{equation}
  P_{\nu} \approx (l_{\rm cool} \Psi_{\rm S}) \times
  J(\nu) \propto \nu^{-\alpha/2} \frac{B^{1+\alpha/2}}{B^{2}+B_{\rm CMB}^{2}}
\end{equation}
where $\Psi_{\rm S}$ is the shock area, and we assume that all
post-shock variables are roughly constant over a distance $l_{\rm
  cool}$. For a given electron population, this simple estimate gives
similar scalings to more careful calculations which integrate over the
cooling layer \citep{2007MNRAS.375...77H}, from the freshly
accelerated to oldest electrons. In particular, we obtain the same
result that cooling steepens the spectral index $\alpha_{\nu}$ from
$(\alpha-1)/2$ to $\alpha/2$. For weak B-fields $B^{2} \ll B^{2}_{\rm
  CMB} = 3.24 (1+z)^{2} \, \mu$G, the luminosity increases with
B-field, $P_{\nu} \propto B^{1+\alpha/2}$. However, this increase
starts to saturate when $B \approx B_{\rm CMB}$; for $B^{2} \gg
B^{2}_{\rm CMB}$, $P_{\nu} \propto B^{\alpha/2-1}$ which is very weak
(for instance, $P_{\nu} \propto B^{1/4}$ for $\alpha=2.5$). This makes
physical sense: when synchrotron emission dominates cooling, then all
of the energy in relativistic electrons is emitted at radio
wavelengths, independent of the B-field.

Is the brightness of observed relics consistent with our calculated
fossil electron population? This question is most accurately answered
with the handful of observations with high spatial resolution or
favorable geometry where the effects of spectral aging can be resolved
or otherwise minimized. We list these in Table
\ref{tab:relic_sample}. The spectral index closest to the shock front,
$\alpha_{\nu} = (\alpha-1)/2$, before cooling steepens the spectrum,
is an accurate measure of shock Mach number. The shock area is
estimated as $\Psi_{\rm S}\sim L^{2}$, where $L$ is largest observed
linear size of relic squared. For simplicity, we assume $B_{2} \sim 5
\mug$ \citep[as for instance estimated for the Sausage relic using spatially
resolved observations of spectral ageing][]{2010Sci...330..347V},
and $kT_{2} \sim 5\,$keV, and only allow variations in $n_{e}$.  Assuming
the median fossil electron spectrum from our simulations, we can match
the observed relic luminosity by reasonable variations\footnote{In
  reality, of course, much of the change comes from the other
  (uncertain) degenerate parameters.} in the electron gas density
$n_{e} \sim 10^{-4} {\rm cm^{-3}}$. Thus, reaccelerated fossil
electrons can clearly produce radio relics of the right luminosities.

\begin{figure}
  \includegraphics[width=1.0\columnwidth]{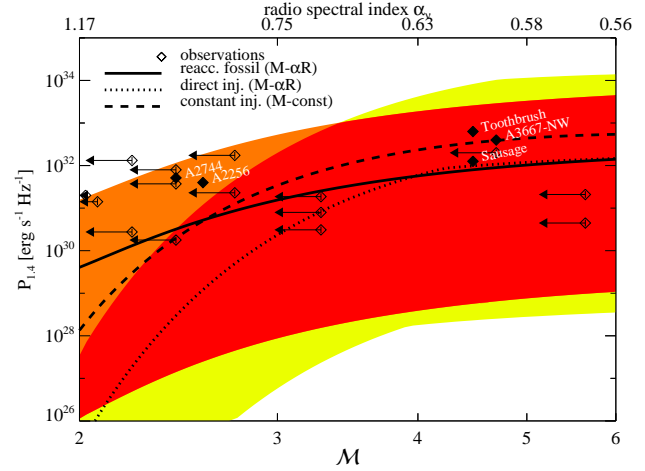}
  \caption{Luminosity from radio relics as a function of Mach number,
    for reacceleration (solid lines), and direct injection with the
    `M-$\alpha$R' (dotted lines) and `M-const' model (dashed
    lines). We also show observations where the intrinsic spectral
    index can be accurately determined, as in Table
    \ref{tab:relic_sample} (filled diamonds; labelled), and those
    where we correct for the effects of cooling (open diamonds; upper
    limits). To bracket the uncertainty in parameters we vary
    parameters (lower limit, fiducial value, upper limit) and color
    code the resulting intervals for our reacceleration scenario of
    fossils (orange), direct injection scenario (yellow) and overlap
    region (red); shock area $\Psi_\rmn{S}=(0.3,2,4)\,\rmn{Mpc}^2$,
    downstream temperature $kT_2=(1,5,10)\,\kev$, electron number
    density $n_\e=(0.1,1,5)\times10^{-4}\,\cm^{-3}$, and $B_2 =
    (1,5,7)\,\mu$G (magnetic field).
    \label{fig:Fluc_Mach}}
\end{figure}

How do the different models compare in their predictions for radio
luminosity as a function of Mach number? In Fig.~\ref{fig:Fluc_Mach},
we contrast the reacceleration model with the direct injection model
(`M-$\alpha$R'), where we adopt fiducial parameters for shock area,
temperature, density and B-field, but also take into account their
possible spread (indicated by the shaded regions). We also show
observations where the intrinsic spectral index can be accurately
determined, as in Table~\ref{tab:relic_sample} (filled diamonds;
labelled; here $\alpha_{\nu} = (\alpha -1)/2$), and those where we
correct for the effects of cooling (open diamonds; upper limits; here
$\alpha_{\nu} = \alpha/2$, derived from the compilation of radio
relics in \cite{2012A&ARv..20...54F}). While discrepancies with the
fiducial curve in the reacceleration model (up to a factor of $\sim
10-30$) lie within the uncertainties, {\it the much larger ($\sim
  100-1000$) discrepancies with the direct injection model means that
  reconciliation is wholly untenable.} The radio flux from
reaccelerated fossil electrons declines smoothly as a function of Mach
number (factor 30 difference between $\mach \sim 2.5$ and $\mach \sim
4.5$) compared to the direct injection scenario, where the flux is
exponentially suppressed as the shock become weaker than $\mach \lsim
4$. This means that as more sensitive radio experiments are built, the
reacceleration model predicts that there will be a substantial
increase in the number of steep spectrum relics, especially since low
$\mach$ shocks are believed to be more abundant that the high $\mach$
shocks. These results for the overwhelming dominance of reaccelerated
fossils at low Mach number mirrors that in Fig.~\ref{fig:f_ratio}, as
it should.

We also show on the same plot results for the constant acceleration
efficiency $\zeta_{\rm e} = 0.01$ (`M-const') model. Note how it does
{\it not} suffer from exponential suppression at low Mach numbers by
fiat, and to some extent mimics the behavior of the reaccelerated
fossil model. This ability to more or less fit the observations is one
reason why it is widely used. However, the physical basis for this
model is somewhat murky. By contrast, fossil electrons do not require
any new (unknown) acceleration physics; in fact, they {\it must}
exist, and the cooling physics which governs their post-injection
evolution is well established.

Although it fares much better than other models, our fiducial
reacceleration model still appears to somewhat underpredict the
luminosities of relics at low Mach number. The observations lie well
within model uncertainties; besides possible variations in area,
temperature, density and B-field in the downstream plasma, we have not
taken into account the important effects of variations in relic size
along the line of sight, viewing geometry, as well as the fact that
the fossil electron abundance can vary by a factor of $\sim 10$
(\S\ref{sect:cluster_vary}). Nonetheless, it is somewhat striking that
at face value the observations do not appear to show any luminosity
trend with Mach number. Several points are worth noting. Firstly, note
that many of the relics have spectral indices that are spatially
unresolved. While we have attempted to correct for the effects of
cooling, depending on viewing geometry and other complications not in
our simple model, spectral steepening due to cooling could be
larger. Thus, the shock Mach number could be underestimated. Secondly,
this could simply be a selection effect, given that flux limits
typically translate to luminosity limits of $P_{1.4} \gsim 10^{30} \,
{\rm erg \, s^{-1} \, Hz^{-1}}$ (except for nearby clusters). Indeed,
counterexamples to the observed bright low Mach number relics also
exist. For instance, in A2146 there is a pair of shocks with Mach
numbers $\mach =2.1,1.6$ which have been unambiguously detected in
X-ray observations, but for which no diffuse radio counterpart has yet
been identified in deep GMRT observations\footnote{Note that the
  absence of radio emission given the flux limits is consistent with
  our model uncertainties.}  \citep{russell11}. Volume-limited samples
or shocks selected in X-ray rather than radio would be required to pin
down selection bias. Thirdly, characterizing shocks by a single Mach
number might be too simplistic; in reality shock fronts are curved
with a spatially varying Mach number \citep{skillman12}. Finally, we
shall see below that the non-linear mapping between $\mach$ and
$P_{1.4}$ imply that despite the predominance of low Mach number
shocks, most relics in fact cluster about a limiting luminosity close
to the asymptotic value of $P_{1.4}$.

\begin{figure}
  \includegraphics[width=1.0\columnwidth]{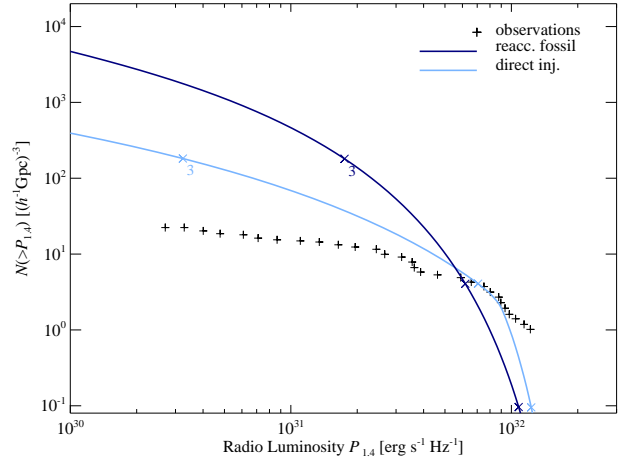}
  \caption{Cumulative number density of radio relics in a $1
    (h^{-1} \, {\rm Gpc})^3$ box. We use the differential shock
    surface model from Kang and Ryu (equation~\ref{eq:KR_shocksurface}),
    with reaccelerated fossil electrons and direct injection
    (`M-$\alpha$R' model) and parameters typical for a relic (see
    Fig.~\ref{fig:Fluc_Mach}). Crosses indicate Mach numbers at unit
    intervals, starting with the numbers on the left. For comparison,
    we show the NVSS relic luminosity function (+). \label{fig:dndl}}
\end{figure}

We now make some approximate calculations for the radio relic
luminosity function. The differential number density of relics as a
function of Mach number can be written as \citep{2003ApJ...593..599R}:
\begin{equation}
  \frac{\dd n}{\dd \mach} = \frac{1}{\Psi_\rmn{S}} \frac{\dd
    S}{\dd \mach}\,,
\end{equation}
where $S$ is the total shock surface area divided by the volume of the
simulation box. Thus, $1/S$ has units of length and can be thought of
as the mean separation between shocks. We assume a typical area for
relic shocks of $\Psi_\rmn{S} = 2\,\rmn{Mpc}^2$. We made a fit for the
Mach number dependent differential shock surface. It is derived from
the shock surface of gas with a pre-shock temperature of $T_{1} \ge
10^{7}$K in a box of size $85 \, h^{-1}$Mpc \citep{kang11}:
\begin{equation}
  \label{eq:KR_shocksurface}
  \frac{\dd S}{\dd \mach} = 10^{-1.015-1.589\mach}\,
  (h^{-1}\,\mpc)^{-1}\,.
\end{equation}
The differential number density of radio relics as a function of
luminosity is:
\begin{equation}
  \frac{\dd n}{\dd P_{1.4}} = \frac{\dd n}{\dd \mach} \frac{\dd
    \mach}{\dd P_{1.4}}\,.
\end{equation}
where ${\dd \mach}/{\dd P_{1.4}}$ depends on the acceleration model;
we use the results of Fig.~\ref{fig:Fluc_Mach} for reacceleration and
direct injection (`M-$\alpha$R'). The total number of radio relics
inside a cosmological box of $(1\,h^{-1}\,\rmn{Gpc})^3$ and with a
luminosity above $P_{1.4}$ is:
\begin{equation}
  N(>P_{1.4}) = (1\,h^{-1}\,\rmn{Gpc})^3 \times \int \dd P_{1.4}
  \frac{\dd n}{\dd P_{1.4}}\,.
\end{equation}
The results are shown in Fig.~\ref{fig:dndl}. Two features stand out:
(1) the reacceleration and direct injection models do not differ
significantly in shape, even at the faint end. (2) The model is
strongly discrepant with observations for low luminosities. Let us
discuss these in turn.

How can we understand the shape of the relic luminosity function?  Its
features can be broadly understood from the results of
Fig.~\ref{fig:Fluc_Mach}. Consider the crosses on the curves for the
\citet{kang11} parametrization in Fig.~\ref{fig:dndl}. These mark
Mach numbers in unit intervals, starting from 2 (3) for reacceleration
(direct injection). The convex shape of the luminosity function is due
to the fact that luminosity falls sharply at low Mach number (even for
the reacceleration model). Thus, a narrow range in Mach number
corresponds to a wide range in luminosity, which `stretches out'
N($>P_{1.4}$) to the left at low luminosity and leads to its flat
slope, since there are relatively few clusters in an interval
$dL$. This amount of `stretching' produced by this non-linear
transformation differs between the two models, as evidenced by the
different Mach number ticks. However, the small difference in faint
end slopes this produces is insufficient to serve as a robust
discriminant between models. The offset on the x-axis, due to
different limiting luminosities, simply reflects assumptions about
asymptotic acceleration efficiencies. The differential luminosity
function is similarly poor at distinguishing between models, where the
main feature is a sharp peak at the limiting luminosity. The bottom
line is that the non-linear mapping between $\mach$ and $P_{1.4}$
causes most luminosities to cluster about a characteristic value,
which mitigates the efficacy of the luminosity function as a
discriminant between models.

The discrepancy of our model predictions with observations is
interesting. Such over-prediction of relics also afflicts most current
models of the relic luminosity function\footnote{As can also be
  inferred from the fact that we obtain roughly similar numbers. For
  instance, the number of relics with a luminosity
  $P_{1.4}>10^{30}\,\rmn{erg}\,\rmn{s}^{-1}\,\rmn{Hz}^{-1}$, within a
  volume of $(500)^{3} (h^{-1} \mpc)^{3}$, is about 1000, consistent
  with what was found in \citet{2012MNRAS.423.2325A} for an
  acceleration efficiency of $\zeta_{\rm e} \sim 0.1$ per cent. For
  LOFAR Tier~1 that has a high sensitivity of about $0.5$~mJy at
  240~MHz \citep{2010arXiv1001.2384M} ($\sim 5\times
  10^{28}\,\rmn{erg}\,\rmn{s}^{-1}$ for an average source distance of
  300~Mpc), we expect about 3000 radio relics could be visible within
  a volume of $(500)^{3} (h^{-1} \mpc)^{3}$, compared with $\sim 2500$
  in \citet{nuza11}.}. Part of the reason may well be that current
surveys are simply highly incomplete, particularly at low
luminosities. Note also that: (1) the $\dd\,S/\dd\,\mach$ function
differs significantly between parametrizations \citep[compare
  e.g.][]{2003ApJ...593..599R,kang11,2012MNRAS.423.2325A}, perhaps due
to differences in the simulation method, shock finding algorithms, and
resolution. Given these differences between simulations, it is perhaps
not surprising that they also disagree with observations. Also note
that our median distribution function is valid only for the cluster
outskirts (which is not singled out by the temperature cut for ${\rm
  d\,S/d}\,\mach$); the distribution function closer in suffers
increased Coulomb cooling and has a lower normalization. (2) We have
assumed fixed values for most relic parameters, whereas in reality
they should have broad distributions, which would produce a tail at
high luminosities and smooth out the steep number density
function. The parameters we adopt are likely biased by selection
effects. For instance, the magnetic field could be highly patchy and
inhomogeneous, and only regions which have a strong pre-existing field
(to be amplified as a shock) are visible as radio relics.

The above issues are well beyond the scope of this paper, though they
deserve careful study in the future. Our main conclusion is that
spectrally resolved observations which constrain Mach numbers can
distinguish between reacceleration and direct injection, while the
relic luminosity function, even with outstanding statistics, cannot.

\section{Discussion: Insights from Heliospheric Observations}
\label{sec:discussion} 
Here, we discuss some uncertainties in our adopted shock physics and
numerical method. In particular, we consider our assumptions that the
acceleration efficiency plummets at weak shocks, and that the shock
surface can be characterized by a single Mach number with little
small-scale spatial or temporal structure. We play particular
attention to insights gained from heliospheric observations.

\subsection{Acceleration Efficiency at Weak Shocks} 
Are physical processes in the solar system relevant to the ICM? At
first blush, the two may appear to differ significantly in their
magnetization. While the plasma in the solar wind is strongly
magnetized with a plasma beta parameter of $\beta=P_{\rmn{th}}/P_B
\sim 0.1-1$ \citep[i.e., the ratio of thermal-to-magnetic
  pressure,][]{2009A&ARv..17..409T}, the bulk of the intra cluster
plasma is only weakly magnetized with $\beta\gtrsim50$ as inferred
from Faraday rotation measure studies \citep[e.g., in
  Coma,][]{bonafede10}. Those, however, are biased towards the denser
core regions and may not be representative of cluster outskirts where
relics form. In fact, the plasma beta parameters at the relics'
position are typically an order of magnitude smaller, $\beta \sim
5$. For instance, comparing the limit on the inverse Compton X-ray
emission with the measured radio synchrotron emission for the
northwest radio relic in A3667 yields a magnetic field estimate of
$B\gtrsim 3\,\mu G$ \citep[or equivalently $\beta\lesssim 3$
  downstream the shock;][]{2010ApJ...715.1143F}. Equating the
advection and synchrotron cooling timescales while deprojecting the
relic width for the northern relic in the cluster CIZA J2242.8+5301
yields $B\simeq 5-7\, \mu G$ \citep[or equivalently
  $\beta\simeq6\pm3$;][]{2010Sci...330..347V,
  2011arXiv1112.3030A,2013MNRAS.tmp..476O}.

What do heliospheric observations tell us about shock acceleration
efficiency as a function of Mach number? Extensive spacecraft
observations in the solar wind have established that electron
acceleration is effectively absent at {\em quasi-parallel weak
  shocks}, i.e. when the upstream magnetic field is approximately
parallel to the shock surface normal (\citealp{1985ApJ...298..676S,
  1989JGR....9410011G, 1992SSRv...59..167K, 2006GeoRL..3324104O}; see
also a recent review by \citealp{2009A&ARv..17..409T}). Electron
heating is mainly dominated by conservation of magnetic moment
\citep{1985GMS....35..195F}. {\em Quasi-perpendicular weak shocks}
(with Alv{\'e}nic Mach numbers $\mach_A \lesssim 3-4$) show electron
acceleration as a result of either shock drift or surfing
acceleration, but diffusive shock acceleration appears unlikely
\citep{1985ApJ...298..676S, 2009A&ARv..17..409T}. This limits the
maximum electron energy to that achievable by shock drift or surfing
acceleration, which also depends on the shock ramp width
\citep{2001GeoRL..28.1367L}. Thus, weak shocks in the solar system do
not result in efficient electron acceleration (through DSA),
consistent with our assumptions for the ICM. One possible
interpretation of this is that in order to solve the injection problem
of electrons for the DSA mechanism, some process operating under
quasi-perpendicular conditions (such as shock drift or surfing
acceleration) would be necessary. By contrast, (rare) {\em strong
  shocks} in the solar system show significant local electron
acceleration up to particle energies of a few MeV \citep[see GEOTAIL
  observations of an interplanetary shock with an Alf\'enic Mach
  number $\mach_\rmn{A}\sim 6$,][]{1999Ap&SS.264..481S}, even under
quasi-parallel magnetic conditions \citep[as demonstrated by Cassini
  observations of Saturn's bow shock with
  $\mach_\rmn{A}\sim100$,][]{2013NatPh...9..164M}. This may hint at
diffusive shock acceleration as the underlying mechanism, in agreement
with predictions by theoretical studies \citep{2010PhRvL.104r1102A}.
  
These observations suggest that magnetic geometry strongly impacts
acceleration efficiencies---at least locally at the shock where
  heliospheric observations are providing measurements. In Appendix
\ref{sect:tests}, we illustrate how a small variation in a geometric
parameter can significantly change the distribution function of fossil
electrons. Nonetheless, we do not propagate uncertainties in magnetic
geometry across our calculations, and only consider DSA as a source of
acceleration. We believe this is justifiable in observed radio relics,
on two grounds: (1) The radio spectral slopes are consistent with
power-law electron spectra due to diffusive shock acceleration. The
high Lorentz factors of $\gamma\sim10^4$ required by the radio data
cannot be achieved by shock drift or surfing acceleration. (2) The
smoothness of the radio intensity and polarization along the shock
surface on scales of $\sim 2$~Mpc argues for a robust acceleration
process, independent of the local field geometry. Nonetheless,
magnetic geometry is an important caveat in considering the ensemble
of possible radio relics: for instance, perhaps only a fraction with
favorable geometry are visible.

\subsection{Shock Surface} 
\label{sect:shock_surface} 
In this paper, we have characterized shocks by a single Mach number
(this is particularly important in \S\ref{sec:flux}, when we compare
our model to observations). Even in the DSA framework, this is an
over-simplification \citep{skillman12}. More generally, the shock
surface is likely to be complex and dynamic. For instance, the
crossing of the termination shock in the solar wind by Voyager 2
revealed the presence of a highly dynamic shock surface. Instead of a
single crossing of a stationary shock, Voyager 2 passed the shock
several times on its ballistic orbit into the heliosheath. The data
reveal a complex, rippled, quasi-perpendicular
supercritical\footnote{A `supercritical' shock is one in which
  gyrating ions represent the primary dissipation process.}
magnetohydrodynamic shock that undergoes reformation on a scale of a
few hours \citep{2008Natur.454...75B}, which corresponds to a
characteristic reformation length scale of $\sim 10^{11}\,\rmn{cm}
\sim 0.01 $~AU (for a solar wind shock velocity of 300 km~s$^{-1}$).
Supercritical shocks are not capable of generating sufficient
dissipation for the retardation, thermalization and entropy increase
associated with a shock transition. When the flow exceeds a critical
Mach number for a given magnetic obliquity, the supercritical shock
starts reflecting particles back upstream. Beyond a second critical
Mach number, whistlers accumulate at the shock front and reform
periodically \citep{2009A&ARv..17..409T}. In addition to these
microscopic shock reformation mechanisms, there should be macroscopic
reformation because of the varying ram pressure and magnetic field
fluctuations in the upstream that cause it to reform on gradient
length scales of the infalling structures, which likely exhibits a
characteristic size spectrum. These are complications well beyond the
scope of this paper, but they illustrate possible complexities
associated with the shock surface.

More mundanely, our results can depend on the numerical method used to
track the injection of CRs. There are known differences in the
morphology of accretion shocks in SPH versus mesh-based finite volume
methods. Entropy profiles show a sharp peak located around $\sim 2-3
\, R_\rmn{vir}$ in mesh-based techniques, while the profiles in SPH
simulations---although similar in shape---are smoothed across a
sizable larger volume, and the volume-weighted Mach number
distribution, particularly for external shocks in low density regions,
have different trends in each code \citep{vazza11}. However, the Mach
number distribution weighted by dissipated energy (which is the most
important quantity here) is reassuringly similar for those different
numerical schemes owing to the conservative nature of the implemented
equations \citep{vazza11}. For the resolution used here, we hence do
not expect dramatic changes, though direct confirmation is of course
preferable.

\section{Conclusions}
\label{sect:conclusions}
In this paper, we consider giant radio relics in galaxy clusters as
the manifestation of reaccelerated fossil relativistic electrons,
rather than direct injection from the thermal pool. This idea is not
new and has already been proposed in the literature several times
\citep[e.g.,]{ensslin98,markevitch05,giacintucci08,kang11,kang12,2012AA...546A.124V}.
What {\it has} been conspicuously missing from the literature is an
actual calculation of the fossil electron distribution function, and
whether it has the correct shape and amplitude to account for the
observations. This paper represents the first attempt to do so. We run
a suite of cosmological simulations where cosmic ray electron
injection and radiative and Coulomb cooling are tracked in
post-processing.

Our principal findings are as follows: 
\begin{itemize}
\item{The fossil CRe population is substantial. Without any
  fine-tuning, it is at the right level to explain observations.}
\item{Reaccelerated fossil electrons are competitive with direct
  injection at high Mach numbers, and overwhelmingly dominate by
  several orders of magnitude at low Mach numbers $\mach \lsim
  3$. Because it is a relative comparison, this conclusion is fairly
  robust to the (potentially uncertain) acceleration model, and
  depends on well-understood structure formation and cooling
  physics. Given that low Mach number shocks are strongly prevalent in
  clusters, we predict that LOFAR should find many more unexpectedly
  bright steep-spectrum radio relics, which cannot be explained by
  direct injection.}
\item{The fossil distribution function has a generic shape with
  distinct regimes, where different cooling processes dominate: (1)
  sub-relativistic Coulomb cooling ($p \lsim 1$), (2) relativistic
  Coulomb cooling ($1 \lsim p \lsim 10$), (3) quasi-adiabatic ($10
  \lsim p \lsim 100$), (4) inverse Compton/synchrotron cooling ($p
  \gsim 100$). Although fossil electrons peak in number density at
  $p\sim 1$, due to the sharp change in the nature of Coulomb cooling,
  electrons at higher energies ($1 \lsim p \lsim 100$ and $10 \lsim p
  \lsim 10^{4}$ for strong and weak shocks respectively) contribute
  most to the observationally relevant portion of the reaccelerated
  spectrum. Cooling times in these regimes are relatively long, and
  thus it is possible to use simulation outputs with relatively poor
  time resolution ($\Delta t =$ 100~Myr, 1~Gyr). }
\item{Since both injection and Coulomb cooling change on the
  (relatively long) dynamical timescale, they can be approximated as
  steady. We compare a self-similar analytic model where this
  approximation is made and find excellent agreement with the
  simulations. This enables extremely rapid estimates. Regime (1), (2)
  \& (4) are in steady-state, while (3) grows monotonically with
  time.}
\item{There can be up to a factor of $\sim 10$ scatter in fossil
  electron abundance, depending on a cluster's merger history; cool
  core clusters have a lower abundance. Spatial variations within a
  cluster can also be significant.}
\end{itemize} 

Given these considerations, we strongly advocate that fossil electrons
be considered a key ingredient in interpreting radio relic
observations, rather than an exotic afterthought. Failure to account
for the fossil electrons will lead to erroneous conclusions about the
nature of particle acceleration at weak shocks. Several extensions
immediately suggest themselves. Seed electrons are needed for
turbulent reacceleration, as in a prominent model for radio halos
\citep{2007MNRAS.378..245B}; we are presently studying if fossils from
structure formation are adequate. Maps, more detailed and realistic
luminosity functions and observational predictions are needed. As
discussed in \S\ref{sect:shock_surface}, confirmation and extension of
these results with a grid code would be most welcome. Our main hope is
to stimulate others to pursue such lines of inquiry.

\section*{Acknowledgments}
We would like to thank our referee Dongsu Ryu for a thoughtful report
and thank Hyesung Kang for useful discussions. We acknowledge NSF
grant 0908480 and NASA grant NNX12AG73G for support. SPO also thanks
the KITP (supported by NSF PHY05-51164), the Aspen Center for Physics
(NSF Grant No. 1066293), UCLA for hospitality, and the Getty Center
for inspiring views, during the completion of this paper. CP
gratefully acknowledges financial support of the Klaus Tschira
Foundation.

\bibliography{bibtex/paper}
\bibliographystyle{mn2e}

\appendix

\section{Cooling}
\label{app:cool}
In this section we explain our implementation of cooling in the
"impulsive" scenario in more detail. These are used to evolve the CRe
population in our simulations with finite time resolution. We develop
analytic expressions which allow very rapid calculations.

\subsection{Coulomb cooling}
We start by deriving the shift in momentum $\pin$ at time $t_i$ to
momentum $\pf$ at time $t_f$ due to Coulomb cooling by integrating
equation~\eqref{eq:d_cool}:
\begin{equation}
  \label{eq:C_cool}
 - \int_{\pin}^{\pf} \frac{\dd p'}{b_\rmn{C}(p')} \approx 
 \int_{t_i}^{t_f}\dd t \, b_\rmn{C}(t) \equiv F(t_f,t_i)\,,
\end{equation}
where $b_\rmn{C}(t) = n_\e(t)/\tilde{n}$, and $\tilde{n}$ an auxiliary
variable introduced to have the correct units. The momentum dependent
part is determined by
\begin{equation}
  - \int \frac{\dd p'}{b_\rmn{C}(p')} = 
  \int \frac{n_\rmn{fix}\,\dd p'}{b_\rmn{C}(p',t)}
\end{equation}
for a fixed electron number density $n_\rmn{fix}=10^{-4}$~cm$^{-3}$,
and where $b_\rmn{C}(p,t)$ is given by equation~\eqref{eq:b_C}. The
integral over time is performed over snapshots that are discrete in
time. Hence, we approximate the time integral for Coulomb cooling
between time $t_i$ and $t_f$ by a discrete sum:
\begin{align}
  \label{eq:gamma_low_sum}
  &F(t_f,t_i) \approx \sum_{j=i+1}^{f} \Delta t_j
  \left[b_\rmn{C}(t_{j-1}) + b_\rmn{C}(t_j)\right]/2\,, \nonumber\\ &
\end{align}
where $j$ denotes the summation index that run over all snapshots
between $t_i$ and $t_f$. Here we have approximated the time integrated
cooling rate between two snapshots by the mean.

First we look at the momentum integral, which we approximate with a
simple fit accurate to 20 per cent within $10^{-2}<p<10^4$, and
$10^{-6}<n_\e<10^{-2}$:
\begin{eqnarray}
  \label{eq:C_cool_p}
  \int \frac{\dd p'}{b_\rmn{C}(p')} &\approx&
  \frac{\kappa\,p^3}{\phi+p^2}\quad \rmn{where}\,,\nonumber
  \\ \kappa&=&7.73\times10^{11}\,\rmn{s}\,,\quad\rmn{and} \quad
  \phi=3.0\,.
\end{eqnarray}
We introduce a characteristic momentum for the Coulomb cooling, $\pC$,
such that
\begin{eqnarray}
  \label{eq:C_low}
  \frac{\kappa\,\pC^3}{\phi+\pC^2} &\equiv& \frac{\kappa\,\pin^3}{\phi+\pin^2} - 
  \frac{\kappa\,\pf^3}{\phi+\pf^2} = F(t_f,t_i)\,,\,\,\rmn{where}\\
  \label{eq:p_C_low}
  \pC &=& \frac{1}{6}\left\{2^{2/3}
  \left(J\,F-2\,F^3+3^{3/2}\sqrt{K\,J}\right)^{1/3}\right. \nonumber\\
  &+& \left.2\,F\left[1 +\frac{2^{1/3}\,F}
{\left(J\,F - 2\,F^3 +3^{3/2}\sqrt{J\,K}\right)^{1/3}}\right]\right\}\,,\nonumber\\
\end{eqnarray}
and
\begin{eqnarray}
  K &=& \phi\,F^2\, \rmn{and} \quad J = 27\,\phi+4\,F^2\,.
\end{eqnarray}
Note that $\pC \sim b_{\rm C}(p,t) \Delta t$ in the relativistic
regime. The shift in momentum is then given by
\begin{eqnarray}
  \label{eq:Delta_C_low}
  \Delta \pC &=& \pf - \pin \\
  \pin &=& \frac{1}{6}\left\{2^{2/3}\left(J'\,F'-2\,F'^3+3^{3/2}\sqrt{K'\,J'}\right)^{1/3} \right.\nonumber\\
  &+& \left. 2\,F'\left[1 +\frac{2^\frac{1}{3}\,F'}{\left(J'\,F' - 2\,F'^3 +3^\frac{3}{2}\sqrt{J'\,K'}\right)^\frac{1}{3}}\right]\right\},\nonumber\\
&&
\end{eqnarray}
and
\begin{eqnarray}
  F' &=& \frac{\pf^3}{\phi+\pf^2}+\frac{\pC^3}{\phi+\pC^2}\, \\ 
  K' &=& \phi\,F'^2\, \rmn{and} \quad J' = 27\,\phi+4\,F'^2\,.  
\label{eq:FKJ} 
\end{eqnarray}

\subsection{Inverse Compton Cooling}
We similarly derive the shift in momentum due to inverse Compton
cooling by integrating equation~\eqref{eq:b_IC_evolu} from the
redshift $z_i=z(t_i)$, where the electrons are injected, to a later
time $z_f=z(t_f)$. We define a characteristic momentum $\pIC$, where:
\begin{equation}
  \label{eq:b_max}
   \frac{1}{\pIC} \equiv \frac{1}{\pf} - \frac{1}{\pin} \approx 
   \frac{b_{\IC,0}}{H_0\,\pf^2}\int_{z_f}^{z_i} \frac{(1+z)^4\,\dd z}{(1+z)\,
\sqrt{\Omega_\rmn{M}(1+z)^3+\Omega_\Lambda}}\,.
\end{equation}
Note that $\pIC \sim 1/(b_{\IC,0}\,\Delta t)$ for relatively short
timescales (where $\Delta t \lesssim 1\,\rmn{Gyr}$, and $b_{\IC,0}$ is
given by equation~\eqref{eq:b_IC}). After the time $\Delta t =
t_f-t_i$ has elapsed, all CRes with a momentum above $\pIC$ have
cooled to lower momentum. The shift in momentum is given by $\Delta
\pIC=\frac{-\pf^2}{\pIC-\pf}$.

\subsection{Cooled Distribution Function}
Given an initial energy $\pin$ of an electron at time $t_i$,
equation~\eqref{eq:d_cool} can be integrated to give the value of
$\pf$ at a later time $t_f$. The differential population density for
relativistic electrons is then given by
\begin{equation}
  \label{eq:f_evolv}
  f_\rmn{inj,CRe}(\pf,t_f,t_i) = f_\rmn{inj,CRe}(\pin,t_i)\,
  \left.\frac{\partial\pin}{\partial\pf}\right|_{t_f}\,,
\end{equation}
where
\begin{equation}
  f_\rmn{inj,CRe}(\pin,t_i) = f_\rmn{inj,CRe}(\pf-\Delta \pIC-\Delta \pC,t_i)\,,
\end{equation}
and
\begin{eqnarray}
&&\left.\frac{\partial\pin}{\partial\pf}\right|_{t_f} = 
\frac{\pIC^2}{\left(\pIC-\pf\right)^2}\,
 -1+\nonumber\\
&&  \left(\frac{\pf}{\pin(\pC)}\right)^2 
\frac{3\,\phi+\pf^2}{\left(\phi+\pf^2\right)^2}  
\frac{\left(\phi+\pin(\pC)^2\right)^2}{3\,\phi+\pin(\pC)^2}\,.
\label{eq:f_evolv_diff}
\end{eqnarray}
The total electron spectrum is derived from the sum of all
individually cooled injected spectra, starting from the time of
injection $t_i$ until a later time $t_f$,
\begin{eqnarray}
  f_\CRe(\pf,t_f) = \sum_j f_\rmn{inj,CRe}(\pf,t_f,t_j)\,.
\end{eqnarray}

\begin{figure}
  \includegraphics[width=1.0\columnwidth]{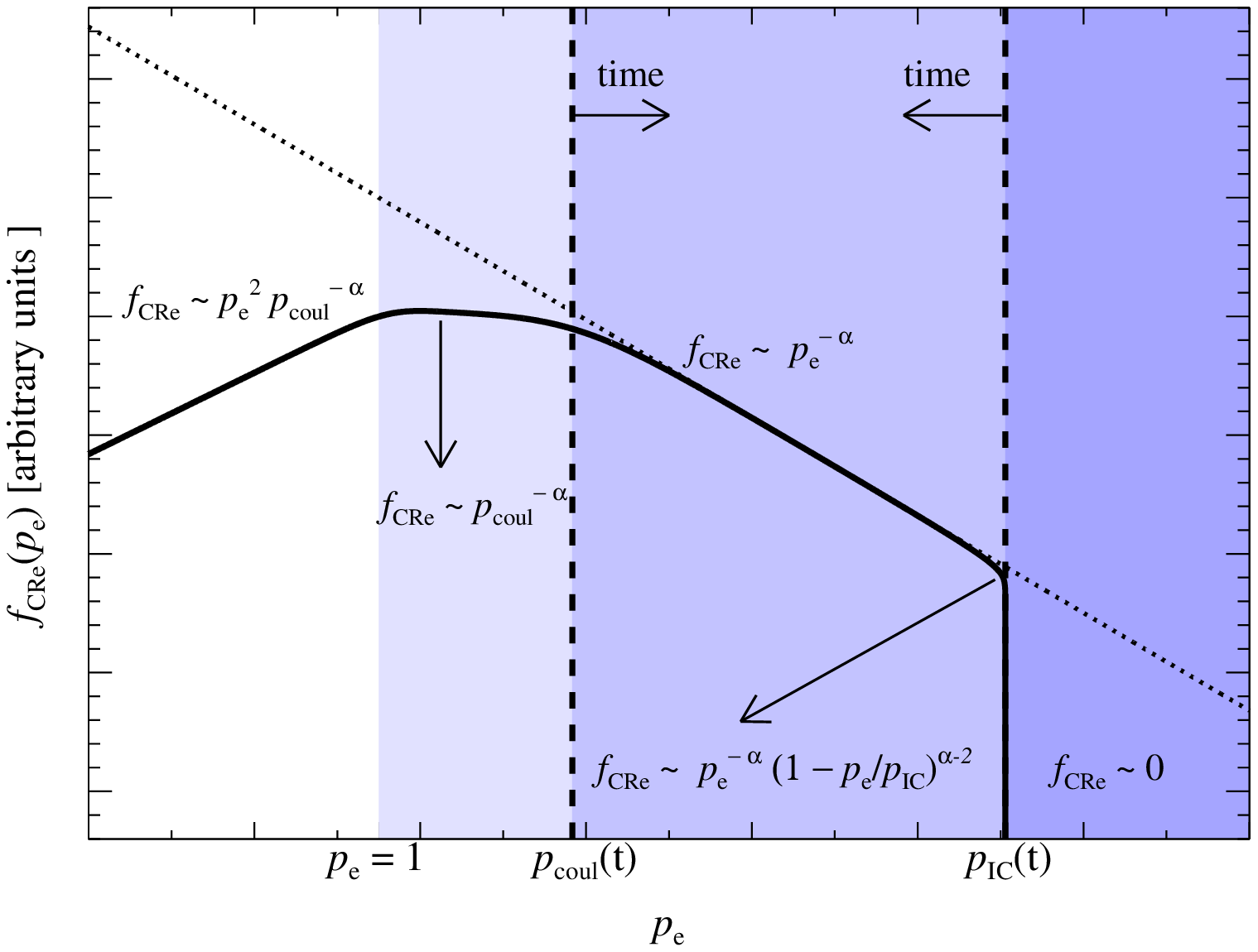}
  \caption{Spectral behavior of the cooled CR electron spectra for
    different momentum regions. There are several distinct momentum
    regimes: (1) the low momentum regime ($p \lesssim 1$) is dominated
    by sub-relativistic Coulomb losses, which cool the CR electrons
    very efficiently, (2) the less efficient relativistic Coulomb
    cooling regime ($1 \lesssim p \lesssim 10$), (3) the adiabatic
    regime ($10 \lesssim p \lesssim 10^2$), which preserves the
    injected distribution function, and where Coulomb and inverse
    Compton/synchrotron losses are less important. (4) the inverse
    Compton/synchrotron cooling regime ($p \gtrsim 10^2$). We assume a
    power-law index of the injected CR electrons of
    $\alpha=2.5$.\label{fig:f_CRe_mom}}
\end{figure}

\section{Dependence on injection parameter}
\label{sect:tests}
\begin{figure}
  \includegraphics[width=0.99\columnwidth]{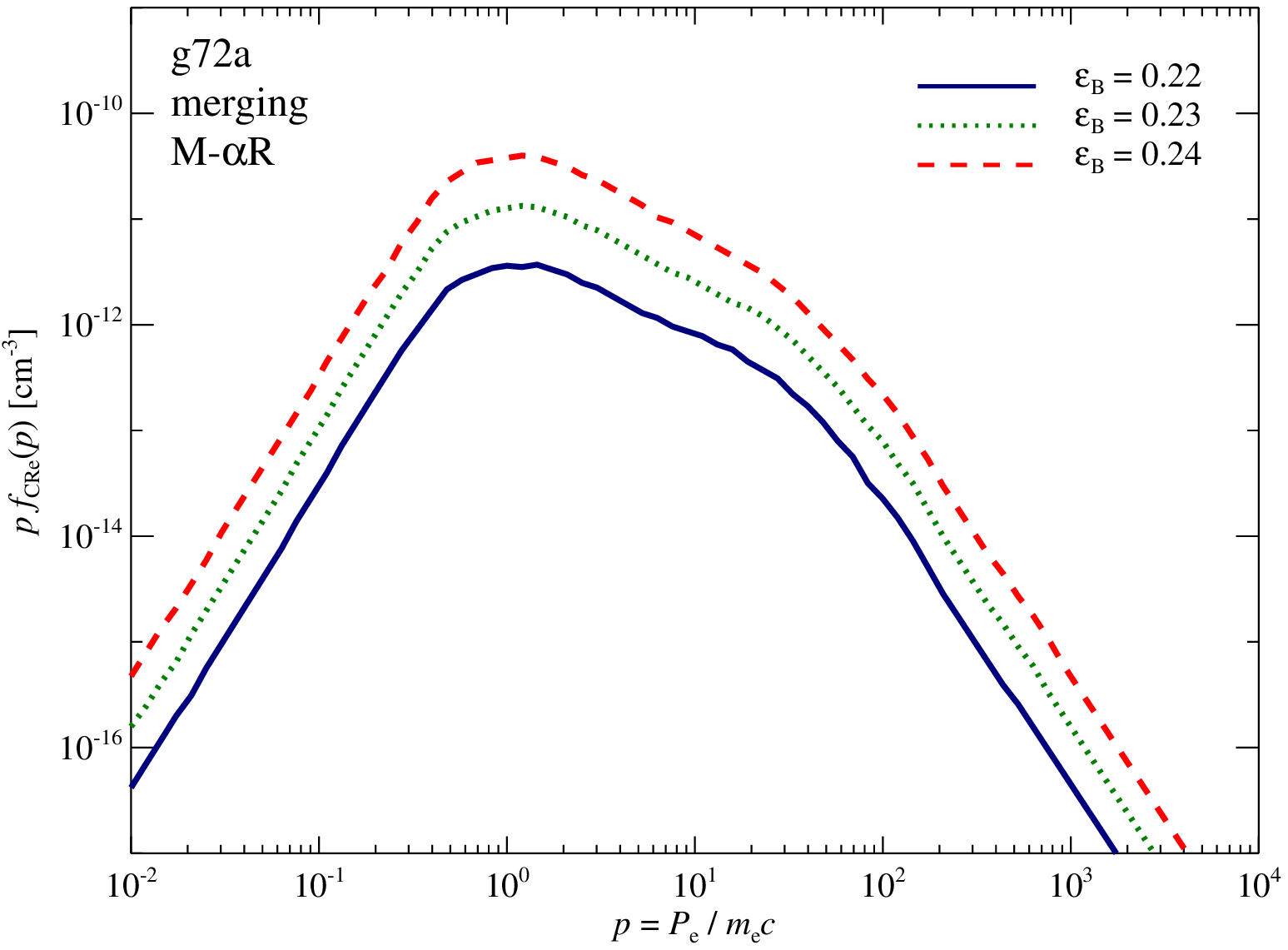}
   \caption{The CR electron dependence on the acceleration parameter
     $\eb= B_0/B_{\perp}$. We show the median fossil CR electron
     spectra with full cooling for a large post-merging cluster at
     $z=0$ in the region between $(0.8-1.0)\,\rvir$. The spectra are
     shown for different values of $\eb$; the value used in the paper
     $\eb=0.23$ (green dotted line), $\eb=0.22$ (blue solid line), and
     $\eb=0.24$ (red dashed line).\label{fig:e_spec_tests}}
\end{figure}

Here we investigate the importance of the parameter that controls the
leakage of the injection process, $\eb= B_0/B_{\perp}$. It is given by
the ratio of the amplitude of the downstream MHD wave turbulence,
$B_0$, to the magnetic field along the shock normal, $B_{\perp}$. The
physical range of this parameter is quite uncertain due to complex
plasma interactions. We adopt a value of $\eb = 0.23$ in the paper,
which corresponds to about a maximum acceleration efficiency of
electrons of $\zeta_{\rm e,inj}=0.01$. In Fig.~\ref{fig:e_spec_tests}
we show how a small variation in this parameter change the
distribution function of the fossil CRes. We find a strong dependence
on this parameter. However, since $\eb$ is essentially degenerate with
$\zeta_{\rm e,inj}$, it cannot vary too much from our assumed value,
as that would conflict with observations of supernova remnants and
radio relics.

\section{Analytic fit to CR Injection Rate}
\label{sect:CRinjection} 
\begin{figure*}
  \begin{minipage}{2.0\columnwidth}
    \includegraphics[width=0.5\columnwidth]{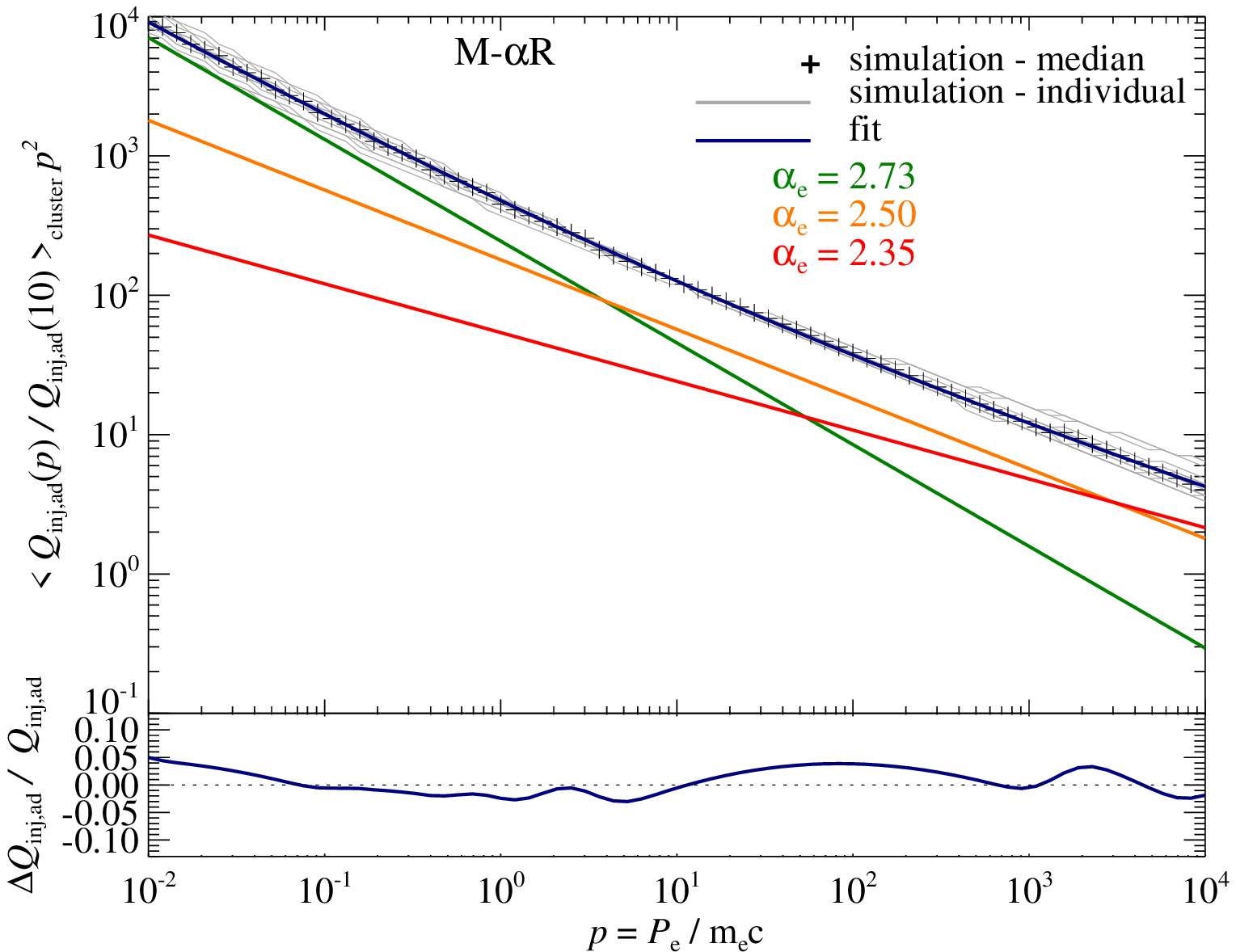}
    \includegraphics[width=0.5\columnwidth]{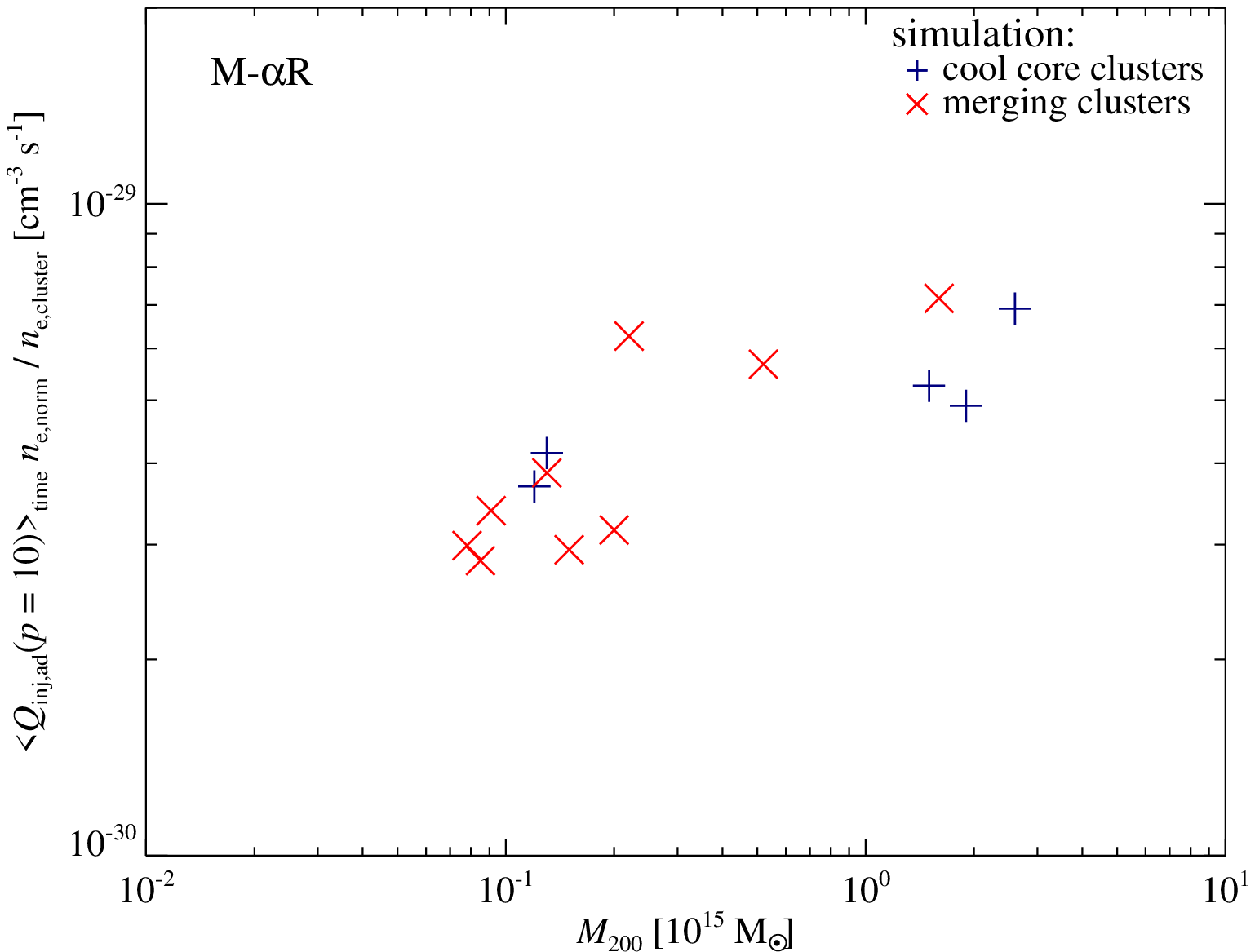}
    \caption{Cosmic ray electron injection rate. We show the median
      injection rate of CR electrons in the cluster region between
      $(0.8-1.0)\,\rvir$ for our sample of 14 simulated galaxy
      clusters. The upper panel in the left figure shows the spectra
      as a function of dimensionless CR electron momenta. The
      injection rate for each cluster is normalized at $p=1$ (grey
      solid lines). The black crosses show the median of the
      normalized injection rate across our cluster sample. The blue
      line shows the best fit triple power-law to the median spectrum,
      where the red, orange and green lines show each of the fitted
      power-laws. The bottom left panel shows the difference between
      the relative fit and the simulation data (blue solid line) which
      amounts to less than 5 percent. The right figure shows the
      injection rate, $<Q_\rmn{inj,ad}(p=1,t_f,t_i)>_\rmn{time}
      (n_\rmn{e,norm}/n_\rmn{e,cluster}(t_f))$, renormalized with the
      electron density $n_\rmn{e,norm}=3\times10^{-5}$~cm$^{-3}$, and
      averaged over time, as a function of galaxy cluster mass $\mvir$
      and cluster morphology. The injection rate for each merging
      cluster (red crosses) and each cool core cluster (blue $+$) are
      shown. Note that our limited statistical sample of clusters
      shows no clear trend with mass.}
    \label{fig:Qinj_fit}
  \end{minipage} 
\end{figure*}
In this Appendix we present an analytic formula for the time-averaged
(over the entire cluster history) CRe injection rate. In the left
panel of Fig.~\ref{fig:Qinj_fit}, we show the injection rate of each
cluster as a function of momentum, where we have normalized all curves
to have the same value at $p=10$ (adiabatic regime). There is
remarkably little difference in the spectral shape, indicating that
all clusters have a similar mix of strong and weak shocks. We use a
triple power-law fit to capture the spectral shape of the median
injection rate over the cluster sample:
\begin{equation}
  Q_\rmn{inj,ad}(p,t_f,t_i)/Q_\rmn{inj,ad}(10,t_f,t_i) = \sum_i A_i
  p^{-a_i}\,,\,\nonumber\\
\end{equation}
where
\begin{align}
  A_0 &= 54.0\,,\quad  &&A_1 = 180.0\,,\quad  &&A_2 = 244.8\,,\quad\rmn{and}\nonumber\\
  \label{eq:Qinj_spectra}
  a_0 &= 2.35\,,    &&a_1= 2.5\,,   &&a_2 = 2.73\,.
\end{align}
This fits to better than 5 per cent over the entire depicted momentum
range.

The injection rate at momentum $p=10$ where the electron number density
is renormalized, $Q_\rmn{inj,ad}(p=10,t_f,t_i)
(n_\rmn{e,norm}/n_\rmn{e,cluster})$, is shown in the right panel of
Fig.~\ref{fig:Qinj_fit}. There is no clear trend with cluster mass or
morphology. Hence, we adopt a median value of
$Q_\rmn{inj,ad}(p=10,t_f,t_i) (n_\rmn{e,norm}/n_\rmn{e,cluster}(t_f)) =
5\times 10^{-30}$~cm$^{-3}$ s$^{-1}$, where we renormalized the density
with $n_\rmn{e,norm} = 3\times10^{-5}$~cm$^{-3}$.

\bsp

\label{lastpage}

\end{document}